\documentclass[conference,10pt]{IEEEtran}

\usepackage[cmex10]{amsmath}
\usepackage{amssymb}
\usepackage{amsfonts}
\usepackage{amsthm}
\usepackage{array}
\usepackage{dsfont}
\usepackage{graphicx}
\usepackage{epstopdf}
\usepackage{cite}
\usepackage{bbm}
\usepackage[hyphens]{url}
\usepackage{cleveref}
\usepackage[caption=false]{subfig}
\usepackage{eurosym}
\usepackage{multirow}
\usepackage{mathtools}
\usepackage{color}
\DeclarePairedDelimiter{\ceil}{\lceil}{\rceil}      
\DeclarePairedDelimiter{\floor}{\lfloor}{\rfloor}   

\allowdisplaybreaks[1]

\makeatletter
\newcommand\footnoteref[1]{\protected@xdef\@thefnmark{\ref{#1}}\@footnotemark}
\makeatother

\IEEEoverridecommandlockouts

\begin{document}

\title{Privacy-Aware Smart Metering:\\ Progress and Challenges}

\author {
  \IEEEauthorblockN{Giulio Giaconi and Deniz G\"{u}nd\"{u}z}

  \IEEEauthorblockA{Imperial College London,  London, UK\\
    {\{g.giaconi, d.gunduz\}}{@imperial.ac.uk}
\and
\IEEEauthorblockN{H. Vincent Poor}
\IEEEauthorblockA{Princeton University, Princeton, NJ, USA \\
poor@princeton.edu}
}
\thanks{The first author gratefully acknowledges the Engineering and Physical Sciences Research Council (EPSRC) of the UK for funding his Ph.D. studies (award reference \#1507704), and BT Research and Innovation for the support provided. This work was also supported in part by the EPSRC through the project COPES (\#173605884), and by the U.S. National Science Foundation under Grants CNS-1702808 and ECCS-1549881.

G. Giaconi and D. G\"{u}nd\"{u}z are with the Department of Electrical and Electronic Engineering, Imperial College London, London,  SW7 2AZ, UK (e-mail: \{g.giaconi, d.gunduz\}@imperial.ac.uk).

H. V. Poor is with the Department of Electrical Engineering, Princeton University, Princeton, NJ 08544 USA (e-mail: poor@princeton.edu).

}}

\maketitle

\begin{abstract}
The next-generation energy network, the so-called \emph{smart grid} (SG), promises a tremendous increase in efficiency, safety and flexibility of managing the electricity grid as compared to the legacy energy network. This is needed today more than ever, as the global energy consumption is growing at an unprecedented rate, and renewable energy sources have to be seamlessly integrated into the grid to assure a sustainable human development. \emph{Smart meters} (SMs) are among the crucial enablers of the SG concept; they supply accurate high-frequency information about users' household energy consumption to a utility provider, which is essential for time of use pricing, rapid fault detection, energy theft prevention, while also providing consumers with more flexibility and control over their consumption. However, highly accurate and granular SM data also poses a threat to consumer privacy as non-intrusive load monitoring techniques enable a malicious attacker to infer many details of a user's private life. This article focuses on privacy-enhancing energy management techniques that provide accurate energy consumption information to the grid operator, without sacrificing consumer privacy. In particular, we focus on techniques that shape and modify the actual user energy consumption by means of physical resources, such as rechargeable batteries, renewable energy sources or demand shaping. A rigorous mathematical analysis of privacy is presented under various physical constraints on the available physical resources. Finally, open questions and challenges that need to be addressed to pave the way to the effective protection of users' privacy in future SGs are presented.
\end{abstract}

\IEEEpeerreviewmaketitle

\section*{Smart Meters for a Smart Grid}

The current energy grid is one of the engineering marvels of the 20\textsuperscript{th} century. However, it has become inadequate to satisfy the steadily growing global electricity demand. In fact, the world energy consumption is predicted to increase $48\%$ from 2012 to 2040 \cite{InternEnergyOutlook2016}, driven by factors such as the growth of world's economy, the rise of the gross domestic product per person, the increase of world's population, an increased penetration of electric vehicles, and a broader mobility revolution \cite{BPEnergyOutlook2016}. Other issues that need to be addressed are the effective integration of renewable energy sources (RESs) and storage capabilities into the grid, the improvement of the grid's environmental sustainability, and the promotion of plug-in hybrid electric vehicles. To address these challenges, a new generation of electricity grid is being engineered, the so-called \textit{smart grid} (SG). SGs are intended to substantially improve energy generation, transmission, distribution, consumption and security, providing improved reliability and quality of the electricity supply, quicker detection of energy outages and theft, better matching of energy supply with demand, and a better environmental sustainability by enabling an easier integration of distributed generation and storage capabilities. The ``smartness" of a SG resides in the advanced metering infrastructure, which enables two-way communication between the utility and its customers, and whose pivotal element in the distribution network is the \textit{smart meter} (SM), the device that monitors a user's electricity consumption in almost real-time. 

\begin{table}[!t]
\caption{List of frequently used acronyms in the paper.}
\centering
\begin{tabular}{ |c|c|c|c| }
\hline
DSO & Distribution system operator\\
\hline
EMP & Energy management policy\\
\hline
EMU & Energy management unit\\
\hline
MDP & Markov decision process\\
\hline
MI & Mutual information\\
\hline
NILM & Non-intrusive load monitoring\\
\hline
RB & Rechargeable battery\\
\hline
RES & Renewable energy source\\
\hline
SG & Smart grid\\
\hline
SM & Smart meter\\
\hline
SMDM & Smart meter data manipulation\\
\hline
SoC & State of charge\\
\hline
ToU & Time of use\\
\hline
TTP & Trusted third party\\
\hline
UDS & User demand shaping\\
\hline
UP & Utility provider\\
\hline
\end{tabular}
\label{tab:nomenclature}
\end{table}

In contrast to legacy grids, in which billing data is gathered at the end of a billing period, SMs send electricity consumption measurements automatically and at a much higher resolution. SMs enable two-way communication with the utility provider (UP), the entity that sells energy to the customers, transmitting a great amount of detailed information. SMs collect and send bidirectional readings of active, reactive and apparent power and energy, i.e., the so-called 4-quadrant metering, that is purchased from the grid, or sold to the grid, if the user produces energy, for example by means of a photovoltaic panel. In the latter case, the user is referred to as a ``prosumer", i.e., a producer and consumer of electricity at the same time, and can be financially rewarded for the energy sold to the grid. SMs also keep track of historical consumption data over the previous days, weeks and months, and provide high-resolution consumption data analytics to the customers to enable them to monitor their energy consumption via an in-home display, web portal or smartphone application, in near real-time. SMs also send alerts about voltage quality measurements, helping UPs fulfill their obligations towards customers concerning energy, power and voltage quality, e.g., in accordance with the European standard EN 50160. Examples of these measurements include the root mean square voltage variations, e.g., voltage dropout, sags and swells, and the total harmonic distortion. Data used for billing, such as current time of use (ToU) tariff, balance and debts, credit and prepayment modes, credit alerts and topping up, is also sent to the UP. SMs can detect if a tampering takes place, and send relevant data about it, along with security credentials for enabling the correct functioning of cryptographic protocols, e.g., hashing, digital signature, and cyclic redundancy check. Finally, SM firmware information and updates are also communicated.

\begin{table}[!t]
\caption{Time resolutions of currently used SMs.}
\centering
\begin{tabular}{ |c|c|c|c| }
\hline
\textbf{Smart Meter Model} & \textbf{Time Resolution} \\
\hline
Itron Centron\footnotemark{} & 1 min \\
\hline
REX2\footnotemark{}  & 5 min \\
\hline
Kamstrup Omnipower\footnotemark{} & 5 min \\
\hline
Enel Open Meter\footnotemark{} & 15 min \\
\hline
\end{tabular}
\label{tab:SMresolution}
\end{table}
\addtocounter{footnote}{-4}
\stepcounter{footnote}\footnotetext{\url{https://www.itron.com/na/technology/product-services-catalog/products/0/7/5/centron}}
\stepcounter{footnote}\footnotetext{\url{https://www.elstersolutions.com/en/product-details-na/826/en/REX2_meter}}
\stepcounter{footnote}\footnotetext{\url{https://www.kamstrup.com/en-us/products-and-solutions/smart-grid/electricity-meters}}
\stepcounter{footnote}\footnotetext{\url{https://www.enel.com/content/dam/enel-com/pressrelease/porting_pressrelease/1666038-1_PDF-1.pdf}}

The increased data resolution is crucial for enabling SG functionalities. Table \ref{tab:SMresolution} shows the smallest time resolution of some SMs currently in use, which is on the order of few minutes. The European Union recommends a time resolution of at least $15$ minutes to allow the new SG functionalities \cite{EUcomm:2011}. For example, the current SM specifications in the UK impose that an SM should send integrated energy readings every $30$ minutes to the UP, while the data sent to a user's in-home display can have a resolution of up to $10$ seconds \cite{decc:2014}. It should be noted that, with the increased adoption of renewable energy generation by the prosumers, the increased penetration of electric vehicles and energy storage technologies, and the diversification of the energy market, it is expected that the SGs will become more volatile, requiring meter readings at a much higher rate in the near future.

SMs provide a wide range of benefits to all the parties in an SG. Thanks to SMs, UPs can gain a better knowledge of their customers' needs, while reducing the cost of meter readings. SMs allow UPs to determine the electricity cost dynamically, as well as to produce more accurate bills, thus reducing customers' complaints and back-office rebilling. The implementation of ToU pricing can incentivize demand response and control customer behavior, while improved demand forecasts and load-shaping techniques can reduce peak electricity demands. Finally, energy theft can be detected more easily and quickly. 

Distribution system operators (DSOs), i.e., the entities that operate the grid, benefit from SMs as well, being able to better monitor and manage the grid. SMs allow DSOs to reduce operational costs and energy losses, and improve grid efficiency and system design, as well as distributed system state estimation and Volt and Var control. Moreover, DSOs are able to better match distributed resources with the ongoing electricity demand and the grid's power delivery capability, thus reducing the need to build new power plants.

Consumers themselves take advantage of SMs to monitor their consumption in near real-time, leading to better consumption awareness and energy usage management. Moreover, consumers receive accurate and timely billing services, with no more estimated bills, and benefit from ToU pricing, by shifting non urgent loads to off-peak price periods.  Microgeneration and energy storage devices can be integrated more easily, and profits from selling the generated excess energy can be collected automatically. Failing or inefficient home appliances, unexpected activity or inactivity, and wasted energy are detected faster and more accurately, and switching between UPs is easier by requesting on-demand readings, which in turn increases the competition among UPs and reduces costs for consumers.

For the above reasons, the installation of SMs is proceeding rapidly, and is attracting massive investments globally. The SM market is expected to grow from an estimated \$12.79 billion in 2017 to \$19.98 billion by 2022, registering a compound annual growth rate of 9.34\% \cite{m&m:2017}. Moreover, the global SM data analytics market, which includes demand response analytics and grid optimization tools, is expected to reach \$4.6 billion by 2022 \cite{tmr:2015}, while the global penetration of SMs is expected to climb from approximately 30\% at the end of 2016 to 53\% by the end of 2025 \cite{navigantResearch:Q3_2016}. These figures show how timely and crucial the research in this field is, and highlight the need to quickly resolve potential obstructions that can threaten the future benefits from this critical technology.



\subsection*{Smart Meter Privacy Risks}

\begin{figure}[t]
\centering
\includegraphics[width=1\columnwidth]{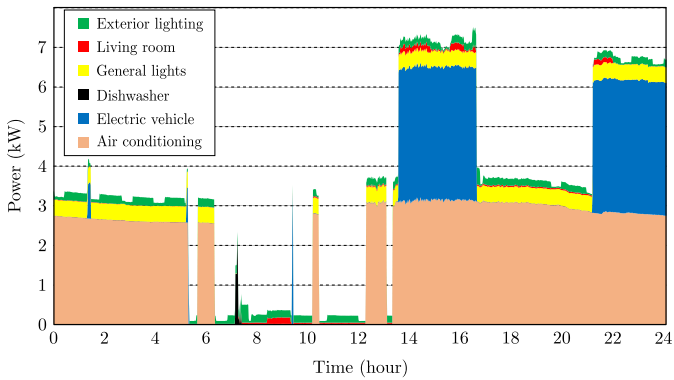}
\caption{An example of a household electricity consumption profile with some appliances highlighted (data retrieved from the Dataport database \cite{pecanstreet}).}
\label{fig:NILMexample}
\end{figure}

\begin{figure}[t]
\centering
\includegraphics[width=1\columnwidth]{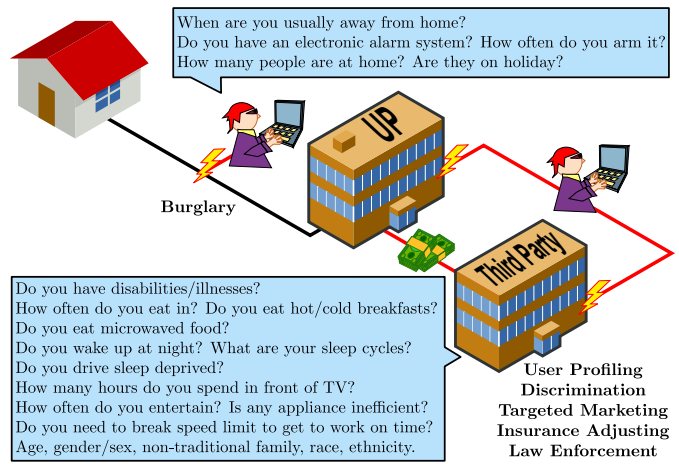}
\caption{Some of the questions an attacker may be able to answer by having access to SM data.}
\label{fig:privacyThreats}
\end{figure}

The SM's ability to monitor a user's electricity consumption in almost real-time entails serious implications about consumer privacy. In fact, by employing non-intrusive load monitoring (NILM) techniques, it is possible to identify the power signatures of specific appliances from the aggregated household SM measurements. NILM techniques date back to the work of George Hart in the 80s, who first proposed a prototype of a NILM device \cite{Hart:1985}. Since then, NILM techniques have improved in different directions, e.g., by assuming either high or low-frequency measurements, by considering known or learned signatures \cite{Hart:1992}, or even by using off-the-shelf statistical methods without any a-priori knowledge of household activities \cite{Molina:2010}. An example of a typical power consumption profile along with some detected appliances is illustrated in Fig. \ref{fig:NILMexample}. As shown in Fig. \ref{fig:privacyThreats}, the UP, a third party that has access to SM data, for example, by buying it from the UP, or a malicious eavesdropper, may gain insights into users' activities and behaviors, and determine, for example, a user's presence at home, her religious beliefs, disabilities, illnesses, and even the TV channel she is watching \cite{Prudenzi:2002,Quinn:2009,Rouf:2012}. Apart from residential users, SM privacy is particularly critical for businesses, e.g., factories and data centers, as their power consumption profile may reveal sensitive information about the state of their businesses to their competitors. SM privacy has attracted significant public attention, and continues to be a topic of heated public and political debate, and even stopped the mandatory SM roll-out plan in the Netherlands in 2009, after a court decided that the forced installation of SMs would violate consumers' right to privacy, and would be in breach of the European Convention of Human Rights \cite{Cuijpers:2012}. Indeed, concerns about consumer privacy threaten the widespread adoption of SMs and can be a major roadblock for this multi-billion dollar industry. 

It is worth pointing out that the privacy problem in SMs is different from the SM data security problem \cite{Loeb:2017}. In the latter there is a sharp distinction between legitimate users and malicious attackers, whereas in the privacy problem in the SM context, any legitimate receiver of data can also be considered malicious. To benefit from the advantages provided by the SG, users need to share some information about their electricity consumption with the UP and DSO. However, by sharing accurate and high-frequency information about their energy consumption, consumers also expose their private lives and behaviors to the UP, which is a fully legitimate user and a potential malicious attacker at the same time. This renders traditional encryption techniques for data privacy ineffective in achieving privacy against the UP, and calls for novel privacy measures and privacy protection techniques.

\section*{Privacy-Enabling Techniques for SMs}

There is a growing literature on privacy-preserving methods for SMs, which can be classified into two main families. The first family, which we call the \textit{SM data manipulation} (SMDM) approach, consists of techniques that process the SM data before reporting it to the UP, while the techniques in the second family, called \textit{user demand shaping} (UDS), aim at modifying the user's actual energy consumption. Considered within the first family are methods such as \textit{data obfuscation}, \textit{data aggregation}, \textit{data anonymization}, and \textit{down-sampling}. Data obfuscation, i.e., the perturbation of metering data by adding noise, is a classical privacy protection method, and has been adapted to SGs in \cite{Kim:2011} and \cite{Bohli:2010}. In \cite{Backes:2014}, \textit{differential privacy}, a well-established concept in the data mining literature, is applied to SMs where noise is added not only to the user's energy consumption, via the RB, but also to the energy used for charging the RB itself to provide differential privacy guarantees. Along these lines, the authors in \cite{Sankar:2013TSG} introduce an information-theoretic framework to study the trade-off between the privacy obtained by altering the SM data and the utility of data for various SG functionalities. Note that, the more noise added to the data, the higher the privacy, but the less relevant and less useful the data is for monitoring and controlling the grid. In \cite{Sankar:2013TSG} an additive distortion measure is considered to model the utility, which allows the characterization of the optimal privacy-vs-utility trade-off in an information-theoretic single-letter form. The data aggregation approach, proposed in \cite{Bohli:2010}, \cite{Garcia:2010} and \cite{Li:2011}, considers sending the aggregate power measurements for a group of households so that the UP is prevented from distinguishing individual consumption patterns. The aggregation can be performed with or without the help of a \textit{trusted third party} (TTP). The data anonymization approach, on the other hand, mainly considers utilizing pseudonyms rather than the real identities of consumers \cite{Petrlic:2010,Efthymiou:2010SGC}. Another method proposed in \cite{Cardenas:2012}, reduces the SMs sampling rate to a level that does not pose any privacy threat. However, the SMDM family suffers from the following shortcomings: 
\begin{itemize}
\item Adding noise to the SM readings causes a mismatch between the reported values and the real energy consumption, which prevents DSOs and UPs from accurately monitoring the grid state, rapidly reacting to outages, energy theft or other problems, and producing accurate and timely billing services. These would significantly limit the benefits of SMs;
\item DSOs, UPs, or more generally any eavesdropper can embed additional sensors right outside a household or a business (street level measurements are already available to the DSOs and UPs) to monitor the energy consumption, without fully relying on SM readings;
\item The anonymization and aggregation techniques that include the presence of a TTP only shift the problem of trust from one entity (UP) to another (TTP).
\end{itemize} 

\begin{figure}
\centering
\includegraphics[width=1\columnwidth]{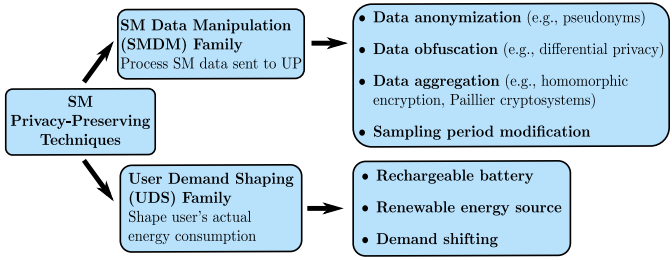}
\caption{Overview of the privacy-enabling approaches for SMs.}
\label{fig:SMprivacyApproaches}
\end{figure}

These issues are avoided by the approaches in the UDS family, which directly modify the actual energy consumption profile of the user, called the \textit{user load}, rather than modifying the data sent to the UP. In this family, the SM reports accurately the energy taken from the grid without any modification; however, this is not the energy that is actually consumed by the appliances. This is achieved by filtering the user's actual electricity consumption via a rechargeable energy storage device, i.e., a rechargeable battery (RB), or by exploiting an RES, which can be used to partially hide the consumer's energy consumption. Examples of RESs include solar panels or micro wind farms. Another technique is to partially shift user's demand. If we denote the energy received from the grid as \textit{grid load}, the idea is to physically differentiate the grid load from the user load. Note that the effect of using an RB or an RES can also be considered as adding noise to the household consumption, but the noise in this case corresponds to a physical variation in the energy received from the grid. Moreover, differently from approaches in the SMDM family, the SM measurements provided by the UDS methods are exact and there is no issue of data mismatch between the SM data and the effective user demand from the grid. Thus, when UDS methods are deployed,  the utility of SMs for the SG is not diminished since the users' energy consumption is neither misreported nor distorted. As a result, while the privacy-vs-utility trade-off is of particular concern for the SMDM techniques, with the UDS techniques smart grid utility is never diminished and other trade-offs are considered instead, such as the \textit{privacy-vs-cost}, or the \textit{privacy-vs-wasted energy}. Fig. \ref{fig:SMprivacyApproaches} shows an overview of the privacy-enabling approaches.


The focus of this article is on UDS techniques, which have been receiving growing attention from the research community in the recent years. The physical resources these techniques rely on, such as RBs or RESs at consumer premises, are already becoming increasingly available, thanks to government incentives and decreasing cost of solar panels and household RBs, as well as the RBs for electric vehicles. Moreover, shaping and filtering users' actual energy consumption by means of physical resources renders any data misreporting or distortion unnecessary, which, thus, do not undermine the utility of the SG concept itself. We will present a signal processing perspective on SM privacy by treating the user load as a stochastic time series, which can be filtered and distorted by using an RB, an RES and/or demand shaping/scheduling. The available energy generated by the RES can also be modeled as a random sequence, whose statistics depend on the energy source (e.g., solar, wind) and the specifications of the renewable energy generator. Additionally, the finite-capacity battery imposes instantaneous limitations on the available energy. We also note that such physical resources can also be used for cost minimization purposes by the users, e.g., by acquiring and storing energy over low-cost periods, and utilizing the stored energy in the RB and the energy generated by an RES over peak-cost periods. Accordingly, we also study the trade-off between privacy and cost, and the minimization of the wasted renewable energy. In the following, we describe and summarize the progress made in recent years towards quantifying the privacy leakage of SMs in a rigorous manner, report the most significant results, and highlight a number of future research directions.

\subsection*{Current Household Batteries, Typical Energy Demands and Renewable Energy Generation}

\begin{table}[!t]
\caption{Specifications of some currently available residential batteries.}
\centering \resizebox{\columnwidth}{!}{
\begin{tabular}{ |c|c|c|c| }
\hline
\textbf{Household RB} & \textbf{Capacity (kWh)} & \begin{tabular}[x]{@{}c@{}} \textbf{RB Charging} \\ \textbf{Peak Power (kW)} \end{tabular} & \begin{tabular}[x]{@{}c@{}} \textbf{RB Discharging} \\ \textbf{Peak Power (kW)} \end{tabular}\\
\hline
Sunverge SIS-6848\footnotemark{} & $7.7$, $11.6$, $15.5$, $19.4$ & $6.4$ & $6$\\
\hline
SonnenBatterie eco\footnotemark{} & $4-16$ & $3-8$ & $3-8$\\
\hline
Tesla Powerwall 2\footnotemark{} & $13.5$ & $5$ & $5$\\
\hline
LG RESU 48V\footnotemark{} & $2.9$, $5.9$, $8.8$ & $3$, $4.2$, $5$ &  $3$,  $4.2$, $5$\\
\hline
\begin{tabular}{@{}c@{}}Panasonic Battery \\ System LJ-SK56A\footnotemark{} \end{tabular}& $5.3$ & $2$ & $2$\\
\hline
\begin{tabular}{@{}c@{}}Powervault \\ G200-LI-2/4/6KWH\footnotemark{} \end{tabular}& $2$, $4$, $6$ & $0.8$, $1.2$ & $0.7$, $1.4$\\
\hline
Orison  Panel\footnotemark{} & $2.2$ & $1.8$ & $1.8$\\
\hline
Simpliphi PHI 3.4-48V\footnotemark{} & $3.4$ & $1.5$ & $1.5$\\
\hline
\end{tabular}}
\label{tab:batteryCapacity}
\end{table}

\begin{table*}[!t]
\caption{Distribution of average household power consumption (resolution refers to the measurement frequency). Values in each column indicate the percentage of time the average consumption falls into the corresponding interval.}
\resizebox{\textwidth}{!}{
\centering
\begin{tabular}{ |c|c|c|c|c|c|c|c|c|c|c| }
\hline
\textbf{Source} & \textbf{Location} & \textbf{Resolution} & \textbf{Time Frame} & \textbf{\# of Houses} & $\mathbf{[0,0.5]}$ \textbf{kW} & $\mathbf{(0.5, 1]}$ \textbf{kW} &  $\mathbf{(1, 2]}$ \textbf{kW} & $\mathbf{(2,3]}$ \textbf{kW} & $\mathbf{(3, 4]}$ \textbf{kW} & $\mathbf{(4,+\infty)}$ \textbf{kW}\\
\hline
\multirow{5}{*}{Dataport \cite{pecanstreet}}  & \multirow{5}{*}{Texas}  & \multirow{5}{*}{$60$ mins} & 01/01/2016 - 31/05/2016    & $512$ & $38$ & $30$   & $20$  & $7$   & $3$ & $2$\\
\cline{4-11}
& &  & 01/01/2015 - 31/12/2015    & $703$ & $36$ & $26$   & $20$  & $9$   & $5$ & $4$\\
\cline{4-11}
& & & 01/01/2014 - 31/12/2014    & $720$ & $39$ & $25$   & $20$  & $8$   & $4$ & $4$\\
\cline{4-11}
& & &01/01/2013 - 31/12/2013    & $419$ & $35$ & $25$   & $21$  & $9$   & $5$ & $5$\\
\cline{4-11}
& & &01/01/2012 - 31/12/2012    & $182$ & $31$ & $26$   & $24$  & $10$  & $5$ & $5$\\
\hline
Intertek \cite{intertek:2012} & UK & $2$ mins & 01/05/2010 - 31/07/2011       & $251$ & $18$ & $24$   & $47$  & $11$  & $0$ & $0$\\
\hline
Dred \cite{dred:2016}& Netherlands & $1$ sec &05/07/2015 - 05/12/2015           & $1$   & $98$ & $1.8$  & $0.4$ & $0$   & $0$ & $0$\\
\hline
Uci \cite{Dua:2017} & France & $1$ min & 16/12/2006 - 26/11/2010   & $1$   & $47$ & $9$    & $28$  & $8$   & $4$ & $2$\\
\hline
\end{tabular}}
\label{tab:SManalitics}
\end{table*}
\begin{table*}[t]
\caption{Distribution of average power generated by residential photovoltaic systems. Values in each column indicate the percentage of time the average generation falls into the corresponding interval.}
\centering
\resizebox{\textwidth}{!}{
\begin{tabular}{ |c|c|c|c|c|c|c|c|c|c|c|c|c|c|c| }
\hline
\textbf{Source} & \textbf{Location}  & \textbf{Resolution}  & \textbf{Time Frame} & \textbf{\# of Houses} & $\mathbf{0}$ \textbf{kW}& $\mathbf{(0,0.5]}$ \textbf{kW} &  $\mathbf{(0.5,1]}$ \textbf{kW}& $\mathbf{(1,2]}$ \textbf{kW}& $\mathbf{(2,3]}$ \textbf{kW}& $\mathbf{(3,4]}$ \textbf{kW}& $\mathbf{(4,+\infty)}$ \textbf{kW}\\
\hline
Dataport \cite{pecanstreet}  & Texas & $60$ min &  01/01/2012 - 31/05/2016 & $351$ & $49$ & $17$ & $7$ & $9$ & $7$ & $6$ & $5$\\
\hline
Microgen \cite{microgen} & UK & $30$ min &  01/01/2015 - 31/12/2015 & $100$ & $51.7$ & $36.4$ & $9.8$ & $2$ & $0.1$ & $0$ & $0$\\
\hline
\end{tabular}
}
\label{tab:photovoltaic}
\end{table*}

\begin{table*}[t]
\caption{Specifications of the solar panels studied in the Microgen \cite{microgen} database. Values in each column indicate the percentage of solar panels that satisfy the corresponding property.}
\centering
\resizebox{\textwidth}{!}{
\begin{tabular}{ |c|c|c|c|c||c|c||c|c|c|c| }
\hline
\multicolumn{5}{|c||}{\textbf{Solar Panel Area ($m^2$)}} & \multicolumn{2}{c||}{\textbf{Solar Panel Cell Type}} & \multicolumn{4}{c|}{\textbf{Nominal Installed Capacity (kWp)}}\\
\hline
$(0,15]$ &  $(15,20]$ & $(20,25]$ & $(25,30]$ & $(30,+\infty)$ & Monocrystalline & Polycrystalline & $(0,2]$ &  $(2,3]$ & $(3,4]$ & $(4,+\infty)$ \\
\hline
$5$ & $35$ & $44$ & $15$ & $1$ & $93$ & $7$ & $4$ & $36$ & $59$ & $1$\\
\hline
\end{tabular}}
\label{tab:photovoltaicSpecifications}
\end{table*}

\addtocounter{footnote}{-8}
\stepcounter{footnote}\footnotetext{\url{http://www.sunverge.com/energy-management/}}
\stepcounter{footnote}\footnotetext{\url{https://sonnen-batterie.com/en-us/sonnenbatterie}}
\stepcounter{footnote}\footnotetext{\url{https://www.tesla.com/powerwall}}
\stepcounter{footnote}\footnotetext{\url{http://www.lgchem.com/global/ess/ess/product-detail-PDEC0001}}
\stepcounter{footnote}\footnotetext{\url{http://www.panasonic.com/au/consumer/energy-solutions/residential-storage-battery-system/lj-sk56a.html}}
\stepcounter{footnote}\footnotetext{\url{http://www.powervault.co.uk/technical/technical-specifications/}}
\stepcounter{footnote}\footnotetext{\url{http://orison.energy/}}
\stepcounter{footnote}\footnotetext{\url{http://simpliphipower.com/product/phi3-4-smart-tech-battery/}}

Table \ref{tab:batteryCapacity} lists the storage capacity and peak power of some of the currently available RBs for residential use. As it can be seen, the capacities are in the range of few kWhs. For example, the peak power that batteries with $4$ kWh capacity can output sustainably is on the order of $1-2$ kW. However, as the typical electricity consumption is very spiky (see Fig. \ref{fig:NILMexample} for an example), current batteries cannot fully hide the spikes in the consumption, due to the charging/discharging peak power constraints. For example, while a $4$ kWh battery can hide a constant consumption of $2$ kW over $2$ hours, it cannot fully hide spikes in the user load of more than $2$ kW. An example of this effect can be noticed in the simulation of Fig. \ref{fig:variance_piece}.

The typical household average power consumption also lies within the range of few kWs, as shown in Table \ref{tab:SManalitics}, where the distribution of the average user power consumption values over different years obtained from various databases is reported with various time resolutions. Analyzing the Dataport database \cite{pecanstreet} we observe that, independently of the period considered, the average user energy demand is less than $2$ kWh for $80-90\%$ of the time. Current batteries charged at full capacity would then be able to satisfy the demand continuously only for a few hours. However, completely covering the consumption over a few hours may come at the expense of revealing the energy consumption fully at future time periods. In fact, once the RB is discharged, it needs to be charged again before being able to hide the user consumption; hence, the use of the RB introduces memory into the system, as decisions taken at a certain time have an impact on the privacy performance at later times. We should also remark that the residential electricity consumption is forecast to increase significantly in the coming years\footnote{https://www.eea.europa.eu/data-and-maps/indicators/total-electricity-consumption-outlook-from-iea/total-electricity-consumption-outlook-from-1}, emphasizing the need to intelligently exploit limited capacity storage devices to hide energy consumption behavior. We also would like to emphasize that the privacy leakage is caused mostly by these spikes, which are typically more informative (e.g., oven, microwave, heater) compared to more regular consumption (e.g., fridge). Moreover, due to electricity price variations users may prefer charging/discharging the battery in certain time periods, which limits the available energy that can be used for privacy. Finally, it is expected that the increasingly wider adoption of electrical vehicles and the mass production and adoption of energy-hungry ``smart devices'' will inevitably increase the typical household electricity consumption, limiting further the rechargeable batteries' capability in fully hiding the user load.

Table \ref{tab:photovoltaic} shows the average power generated via a solar panel, which is the most common residential RES. Locations, technology, as well as the inclination and size of the solar panel affect the generated power, as shown in Table \ref{tab:photovoltaicSpecifications} for one of the databases considered, where kWp denotes the kilowatt peak, i.e., the output power achieved by a panel under full solar radiation. As expected, around $50\%$ of the time, i.e., at night, no energy is generated at all, while there are differences in the distribution of the average values for the two databases considered, due to the different locations. Comparing these values with those in Table \ref{tab:batteryCapacity}, we note that the battery capacities are large enough to store many hours of average solar energy generated by the solar panels most of the time.

\section*{A Signal-Processing Perspective on SM Privacy}

\begin{figure}
\centering
\includegraphics[width=1\columnwidth]{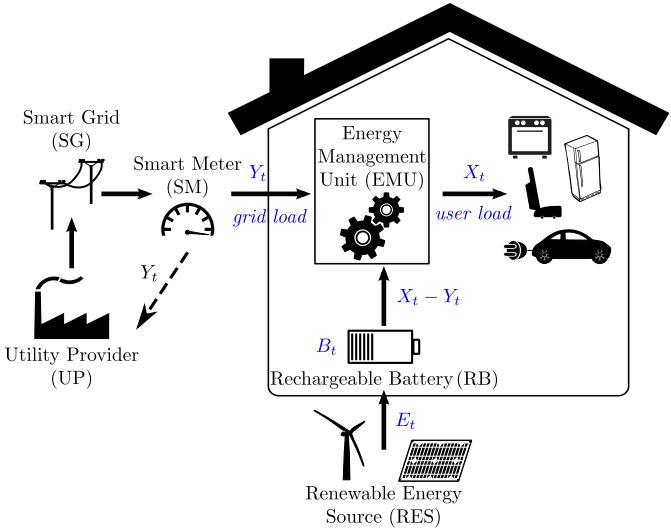}
\caption{System model. $X_t$, $Y_t$, $E_t$ and $B_t$ denote the consumer's power demand, i.e., the \textit{user load}, the SM readings, i.e., the \textit{grid load}, the power produced by the RES, and the battery state of charge at time $t$, respectively. The dashed line represents the meter readings being reported to the UP.}
\label{fig:systemModel}
\end{figure}

The generic discrete-time SM system model is depicted in Fig. \ref{fig:systemModel}. In this model, each time slot is normalized to unit time; therefore, power and energy values within a time slot are used interchangeably. $X_t \in \mathcal{X}$ denotes the total power demanded by the appliances in the household in time slot $t$, i.e., the \textit{user load}, where $\mathcal{X}$ denotes the user load alphabet, i.e., the set of values that $X_t$ can assume. The sequence $\{X_t\}$ denotes the user's private information, which needs to be protected. $Y_t \in \mathcal{Y}$ is the power received from the grid in time slot $t$, i.e., \textit{the grid load}, which is measured and reported to the UP by the SM, while $\mathcal{Y}$ denotes the grid load alphabet. We assume that the user and grid load powers remain constant within a time slot. This can be considered as a discrete-time linear approximation to a continuous load profile in practice. This approximation can be made as accurate as desired by reducing the time slot duration.

In current systems, where no energy manipulation is employed, $Y_t=X_t, \forall t$; that is, the actual energy consumption of the appliances is reported to the UP by the SM. Instead, we will assume that an RB and an RES are available to the user to physically distort the energy consumption; so that what the user receives from the grid, $Y_t$, does not reveal too much information about the energy used by the appliances, $X_t$. We remark here that the time slots in our model correspond to time instants when the electricity is actually requested by the user and drawn from the grid, rather than the typically longer sampling interval used for sending SM measurements to the UP. In fact, we assume that the SM measures and records the output power values at each time slot; this is because our aim is to protect consumers' privacy not only from the UP, but also from the DSO or any other attacker that may deploy a sensor on the consumer's power line, recording the electricity consumption in almost real-time. The state of charge (SoC) of the RB, i.e., the amount of energy stored in the RB, at time $t$, is $B_t \in [0,B_{\max}]$, where $B_{\max}$ denotes the maximum battery capacity. $X_t-Y_t$ denotes the power taken from the RB, and the battery charging and discharging processes are often constrained by the so-called charging and discharging power constraints $\hat{P}_c$ and $\hat{P}_d$, respectively, i.e., $-\hat{P}_c \leq X_t - Y_t \leq \hat{P}_d$, $\forall t$. There is typically a constraint on the average energy that can be retrieved from an RB as well, imposed by an average power constraint $\bar{P}$, i.e., $\mathbb{E}\big[\frac{1}{n}\sum_{t=1}^n (X_t - Y_t)\big] \leq \bar{P}$. Losses in the battery charging and discharging processes may also be taken into account to model a more realistic energy management system. The renewable energy generated at time $t$ by the RES is denoted by $E_t\in \mathcal{E}$, where $\mathcal{E}=[0,E_{\max}]$. RBs and RESs are expensive facilities, and installation and operation costs can be reduced if they are shared by multiple users, e.g., users within the same neighborhood or block of flats. Moreover, sharing these resources allows the centralized management of the energy system, which also leads to a more efficient use of the available resources. The renewable energy can be stored in the RB, or used immediately, so that a user can:
\begin{itemize}
\item \textit{increase privacy}, by avoiding to report her actual power consumption to the UP; 
\item \textit{decrease electricity costs}, by purchasing and storing electricity from the grid when it is cheaper, and use it to satisfy future demand, or even sell it back to the UP when the price increases;
\item \textit{increase energy efficiency}, by reducing the waste of generated renewable energy when it is not needed, and when it is not profitable to sell it to the UP.
\end{itemize}
The random processes $X$ and $E$ are often modeled as Markov processes, or as sequences of independent and identically distributed (i.i.d.) random variables. Although the UP typically does not know the instantaneous realizations of these processes, it may well know their statistics. In some cases, the UP may know the realizations of the renewable energy process $E$, for example, if it has access to additional information from sensors deployed near the household that measure different parameters, e.g., the solar or the wind power intensity, and if it knows the specifications of the user's renewable energy generator, e.g., model and size of the solar panel.

Given the above definitions, the battery SoC update can be expressed as \begin{align}\label{battery_constraint}
	B_{t+1} = \min \Big\{B_{t} + E_t - (X_t - Y_t), B_{\max} \Big\}.
\end{align}

Sometimes, the user load does not need to be satisfied immediately in its entirety. In fact, the user load can be further classified into demand that must be met immediately, e.g., lighting or cooking, and demand that can be satisfied at a later time, the so-called \textit{elastic demand}, e.g., electric vehicle charging, dishwasher or clothes washer-dryer. For the latter demand, the user's only concern is that a certain task needs to be finished by a certain deadline, e.g., her electric car must be fully charged by 8 a.m., and it is not of interest at what exact time the consumption takes place. This flexibility allows the consumer to employ \textit{demand response}  to increase her privacy as well as to lower the energy cost.

The electricity unit cost at time $t$, denoted by $C_t$, can be modeled as a random variable, or in accordance with a specific ToU tariff. The cost incurred by a user to purchase $Y_t$ units of power over a time interval of $\tau_t$ at the price of $C_t$ is thus given by $\tau_t Y_t C_t$. When the presence of an RES is considered, the prosumer may be able to sell part of the energy generated to the grid to further improve her privacy and to minimize the energy cost. If this occurs, the \textit{net metering} approach is typically considered, i.e., the utilities purchase consumer-generated electricity at the current retail electricity rate. The battery wear and tear due to charging and discharging the RB can also be taken into account and modeled as an additional cost \cite{Yang:2015TSG}.

\subsection*{The Energy Management Policy (EMP)} 

The energy management unit (EMU) is the intelligence of the system, located at the user's premises, where the SM privacy-preservation and cost-optimization algorithms are physically implemented. The EMP, implemented by the EMU,  determines at any time $t$ the amount of energy that should be drawn from the grid and the RB, given the previous values of the user load $X^t$, renewable energy $E^{t}$, battery SoC $B^t$, and grid load $Y^{t-1}$, i.e.,
\begin{equation}\label{eq:startingPolicy}
f_t: \mathcal{X}^t \times \mathcal{E}^t \times \mathcal{B}^t \times \mathcal{Y}^{t-1} \rightarrow \mathcal{Y}, \qquad \forall t,
\end{equation}
where $f \in \mathcal{F}$, and $\mathcal{F}$ denotes the set of feasible policies, i.e., policies that produce grid load values that satisfy the RB and RES constraints at any time, as well as the battery update equation in (\ref{battery_constraint}). The optimal policy is chosen to minimize the long-term information leakage about a consumer's electricity consumption, possibly along with other criteria, such as the minimization of electricity cost or wasted energy. The EMP prevents outages, and typically it is not allowed to draw more energy from the grid to be wasted simply for the sake of increased privacy. 

The policy $f_t$ in (\ref{eq:startingPolicy}) corresponds to an \textit{online energy management policy}, i.e., a policy in which the action taken by the EMU at any time slot depends only on the information available causally right up to that time. Alternatively, in an \textit{offline optimization} framework, the policy takes actions based also on future information about the system state, i.e., user load and RES energy generation, in a non-causal fashion. In the SM privacy literature, both offline and online SM privacy-preserving algorithms have been considered. Online algorithms are more realistic and relevant for real-world applications; however, offline algorithms may lead to interesting intuitions or bounds on the performance. Moreover, non-causal knowledge of the electricity price process is a realistic assumption in today's energy networks; and even the non-causal knowledge of power consumption may be valid for certain appliances, such as refrigerators, boilers, heating and electric vehicles, whose energy consumption can be accurately predicted over certain finite time frames.

In the following, we will first present an overview of some heuristic solutions to the SM privacy problem. Next, we will describe the more rigorous and mathematically involved techniques, which are aimed at finding the relevant fundamental bounds and trade-offs.



\section*{A Heuristic Privacy Measure: Variations in the Grid Load Profile}

As in many other problems involving privacy, a wide consensus over the best privacy measure for SMs has not been reached yet, and a number of privacy measures have been proposed in the literature, each with its own benefits and limitations. Although it is clear that privacy is achieved when the UP cannot infer a user's behavior on the basis of SM measurements, it is challenging to define a corresponding mathematical measure that is independent of the particular detection technique employed by the attacker.

\subsection*{Grid Load Variance as a Privacy Measure}

One can argue that privacy in SMs can be ensured by opportunely charging and discharging the RB so that the grid load is always constant. In fact, the differences in consecutive load measurements, $y_t-y_{t-1}$, are indicative of the appliances' switch-on/off events, the so-called \textit{features}, and are typically exploited by the existing NILM algorithms. Ideally, a completely ``flat" grid load profile would not reveal any feature, and would only leak a user's long term average power consumption. However, this would require a very large battery capacity and/or a powerful RES. Alternatively, the level of privacy can be measured by the ``distance'' of the grid load from a completely flat \textit{target load profile}, based on the intuition that the smaller the distance, the higher the level of privacy achieved \cite{Tan:2017TIFS}. Accordingly, privacy can be defined as the grid load variance around a prefixed target load profile $W$, i.e.,
\begin{equation}\label{eq:loadVarianceExp}
\mathcal{V}_n \triangleq \frac{1}{n} \sum_{t=1}^{n} \mathbb{E}\Big[ (Y_t-W)^2\Big],
\end{equation}
where the expectation is over $X_t$ and $Y_t$, and $W = \mathbb{E}[X]$ typically.

Another important concern for consumers is their energy cost. With the integration of unreliable RESs into the grid, it is expected that the unit cost of energy from different UPs will fluctuate in time. RBs for residential use provide flexibility to the consumers as they can buy and store energy during low-cost periods to be used during peak-price periods. The impact of RBs in reducing the cost of energy to the consumers have been extensively studied in the literature \cite{Imp:2016}. Note, however, that the operation of the EMU in order to minimize the energy cost does not necessarily align with the goal of minimizing privacy leakage. Therefore, it is essential to jointly optimize the electricity cost and the user privacy. If the cost of energy and battery wear and tear are considered, the overall optimization problem becomes:
\begin{equation}\label{eq:P1}
\min \frac{1}{n} \sum_{t=1}^n \mathbb{E}\Big[C_t Y_t +\mathbbm{1}_B(t) C_B + \alpha(Y_t - W)^2 \Big],
\end{equation}
where $\mathbbm{1}_B(t) = 1$ if the battery is charging/discharging at time $t$, and $0$ otherwise; $C_B $ is the battery operating cost due to the battery wear and tear caused by charging and discharging the RB; and $\alpha$ strikes the trade-off between privacy and cost. The expectation in (\ref{eq:P1}) is over the probability distributions of all the involved random variables, i.e., $X_t$, $Y_t$, and $C_t$.

\begin{figure}[!t]
\centering
\includegraphics[width=0.9\columnwidth]{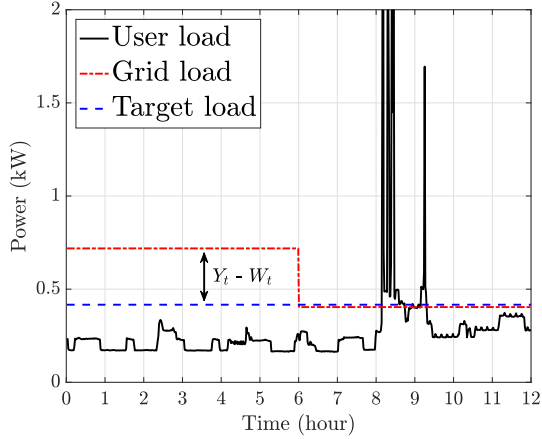}
\caption{Example of the user load, grid load, and constant target load profiles, where the ``distance'' $Y_t-W_t$ is highlighted. The aim of the algorithms presented in this section is to minimize the average squared distance.}
\label{fig:variance}
\end{figure}

If $W_t = \mathbb{E}[X]$, $\forall t$, the EMU tries to achieve a flat grid load profile around the average user energy consumption with as little deviations as possible. This scenario is illustrated in Fig. \ref{fig:variance}, where the straight blue line is the fixed target consumption profile $W_t$, and the red line indicates the achieved grid load profile $Y_t$.  For i.i.d. $X$ and $C$ processes, an online EMP can be obtained using Lyapunov optimization \cite{Yang:2015TSG}. The online control algorithm can be formulated as a Lyapunov function with a perturbed weight and the \textit{drift-plus-penalty} framework is adopted, which is typically used for stabilizing a queuing network, by minimizing the so-called \textit{drift}, while minimizing at the same time a \textit{penalty} function. Here, the penalty is represented by the optimization target, while the \textit{Lyapunov drift} is defined as the difference of the SoC of the RB at successive time instants. Authors in \cite{Yang:2015TSG} show that this approach leads to a mixed-integer nonlinear program, which they solve by decomposing it into multiple cases, and by finding a closed-form solution to each of them.


This problem can also be studied in an offline framework, by assuming the future user demand profile can be accurately estimated for a certain time horizon, and the energy cost is known in advance. When privacy and cost of energy are jointly optimized over a certain time horizon, one can characterize the points on the Pareto boundary of the convex region formed by all the cost and privacy leakage pairs, by solving the following convex optimization problem \cite{Tan:2017TIFS}:
\begin{equation}\label{eq:Tan_2017_JIFS_target}
\min_{Y_t \geq 0} \sum_{t=1}^n \bigg[(1-\alpha) Y_t C_t + \alpha  (Y_t-W)^2\bigg].
\end{equation}
It is shown in \cite{Tan:2017TIFS} that the optimal offline solution has a \textit{water-filling} interpretation. However, differently from the classical water-filling algorithm, which appears as the solution of the power allocation problem across parallel Gaussian channels under a total power constraint, here the \textit{water level} is not constant, and changes across time because of the instantaneous power constraints.

\begin{figure}[!t]
\centering
\includegraphics[width=0.9\columnwidth]{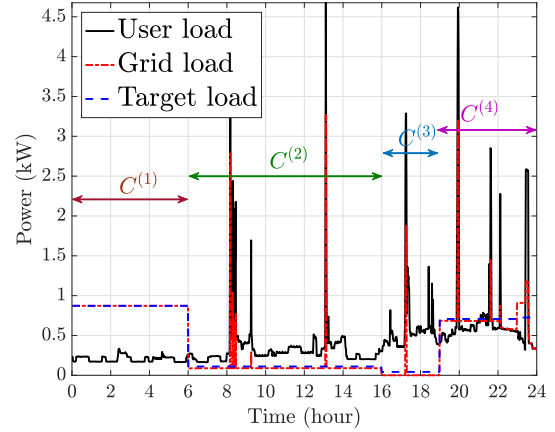}
\caption{Examples of user load, grid load, and target load over the course of a day when a piecewise target profile is considered. The price periods are highlighted by arrows of different colors. Note that the target assumes a different constant value for each price period.}
\label{fig:variance_piece}
\end{figure}

\begin{figure}[!t]
\centering
\includegraphics[width=0.9\columnwidth]{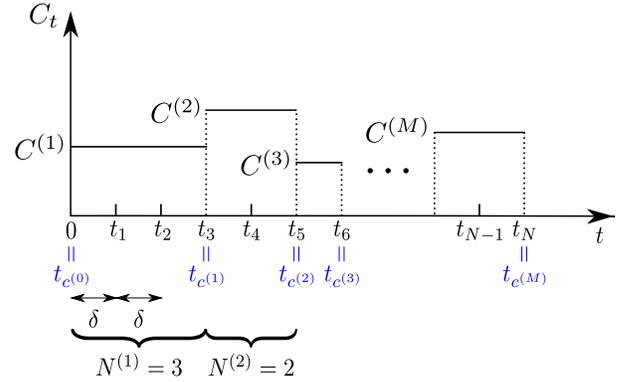}
\caption{ToU tariff and timing convention used for a piecewise target profile \cite{Giaconi:2017SGC}. $t_{c^{(i)}}$, for $i=1,\ldots,M$, are the time instants at which the price of energy changes, and $t_{c^{(0)}}=0$.}
\label{fig:energyCost}
\end{figure}

\begin{figure}[!t]
\centering
\includegraphics[width=0.9\columnwidth]{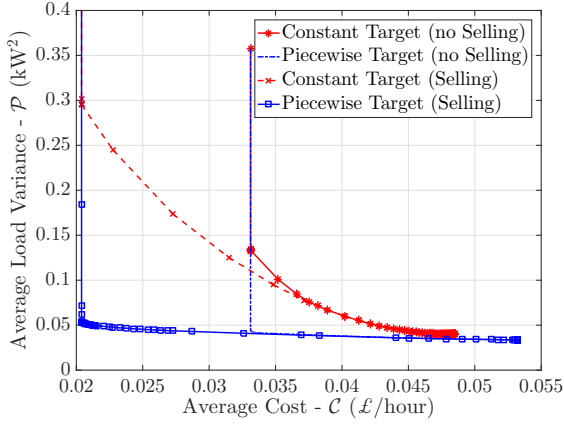}
\caption{Privacy-vs-cost trade-off when using a Powervault G200-LI-4KWH battery for the strategies in \cite{Tan:2017TIFS} and \cite{Giaconi:2017SGC}.}
\label{fig:tradeoff}
\end{figure}

A completely flat consumption profile may not be feasible, or even desirable, for example if the cost varies greatly during the system operation due to ToU tariffs. Thus, it is reasonable to assume that a user requests more energy during off-peak price periods as compared to peak price periods; and hence, allows a piecewise constant target load \cite{Giaconi:2017SGC}. An example of the application of this strategy is shown in Fig. \ref{fig:variance_piece}, applied to real power consumption data from the UK-Dale dataset \cite{UKDALE}. The optimization problem (\ref{eq:Tan_2017_JIFS_target}) becomes
\begin{equation}\label{eq:Giaconi_2017_SGC_target}
\min_{Y_t,W^{(i)}}  \frac{1}{N} \sum_{i=1}^{M} \Bigg[ \sum_{t=t_{c^{(i-1)}}}^{t_{c^{(i)}}} (1-\alpha) Y_t C^{(i)} + \alpha (Y_t - W^{(i)})^2 \Bigg],
\end{equation}
where $C^{(i)}$ and $W^{(i)}$ are the cost of the energy purchased from the UP and the target profile during the $i$-th price period, respectively, where $1 \leq i \leq M$, $M$ is the total number of price periods during time $T$, and the $i$-th price period spans from time slot $t_{c^{(i-1)}}$ to $t_{c^{(i)}}$. Fig. \ref{fig:energyCost} depicts the timing convention considered in this scenario. Energy can be sold to the UP to further improve the privacy-vs-cost trade-off, as assumed in \cite{Giaconi:2017SGC}. Considering a piecewise target profile improves the overall privacy-vs-cost trade-off compared to a constant target profile, as shown in Fig. \ref{fig:tradeoff} for a  Powervault G200-LI-4KWH RB when using power consumption data from \cite{UKDALE}.  

A possible extension of the latter work is considering the multi-user scenario, where, in principle, each user can fix its own target profile. As long as the target profile does not depend on the user's energy consumption profile, the UP does not receive much information about user's activities. On the other hand, the UP can implicitly incentivize users to choose different target profiles by setting different ToU prices for different consumers. Since consumers will tend to buy more energy when it is cheaper, each of the users in the neighborhood will shift her load to a different time slot, also balancing the total load on the grid.

\subsection*{Markov Decision Process (MDP) Formulation}

In the online optimization framework, where the user load and the energy generated by the RES can be modeled as Markov processes (or, as an i.i.d. sequence as a special case), the SM privacy problem can be cast as an MDP. An MDP is a discrete-time state-transition system, which is formally characterized by a \textit{state space}, an \textit{action space}, which includes the possible actions that can be taken by the decision maker at each state, the \textit{transition probabilities} from the current state to the next state, which describe the dynamics of the system, and the \textit{reward} (or, inversely, the \textit{cost}) process, which indicates the reward received (or, cost incurred) by the decision maker by taking a particular action in a particular state. The goal of an MDP is to find the optimal policy that minimizes the average (or, discounted) cost either by a specified time in the future, i.e., by considering the so-called \textit{finite horizon} setting, or over an indefinite time period, by considering the so-called \textit{infinite horizon} setting. To solve the corresponding MDP the optimal Bellman optimality equations should be formulated \cite{Bertsekas:2007}, which can be solved to obtain the optimal policy at each state and time instant. The problem can be solved numerically for the finite horizon setting, while the value iteration algorithm can be employed to obtain the optimal stationary policy in the infinite horizon scenario.

In the SM problem, the state at any time $t$ is typically represented by a combination of the current battery SoC $B_t$, user demand $X_t$, and renewable energy $E_t$; the action, performed by the EMU, is represented by the current grid load and the energy used from the RB and the RES. State transitions are modelled by the battery update equation, which is typically assumed to be deterministic, as well as transitions in the user demand and renewable generation states, which typically do not depend on user's actions. The cost function is the privacy loss that is experienced when moving from one SoC to another by following a certain action. However, to consider the privacy as the cost function in an MDP, it is necessary to formulate the privacy leakage in an additive form across time, so that the total loss of privacy over multiple time slots is given by the summation of the privacy leakage at different time slots. This may be challenging depending on the privacy measure employed. For example, measuring privacy via the squared distance of the grid load from a constant target profile has a straightforward additive formulation, while the same does not hold when privacy is measured by the mutual information (MI) between the user and grid load sequences. This is because the MI takes into account the dependence between the realization of the user load at time $t$, $X_t$, and the current, past and future realizations of $Y$, $Y_1, \ldots, Y_t, \ldots$. 

When the state and action spaces are continuous, it is necessary to discretize them to solve the problem numerically. The accuracy of the numerical solution can be improved by decreasing the discretization step size, at the expense of significantly higher computational complexity. When the dimensions of the state and action spaces render numerical evaluation of the optimal policy unfeasible, one can resort to suboptimal solutions that are easier to optimize and compute numerically, yet may provide near optimal performance or interesting intuitions. Also, when the information theoretic privacy measures are used, it may be possible to simplify the infinite-horizon optimization problem, and write it in a single-letter form. We will provide further insights into this below.

The SM problem is cast as an MDP in \cite{Sun:2018IOTJ}, where the loss of privacy is measured by the fluctuations of the grid load around a constant target load, and the joint optimization of privacy and cost is studied. The optimal privacy-preserving policies are characterized by minimizing the expected total cost. Denote by $u_t$ the action at time $t$. To solve the MDP the transition probabilities $p(X_t|X_{t-1})$ and $p(B_t|B_{t-1},u_t)$ need to be known; however, this is normally not the case, as the user load and the energy storage usage are typically non-stationary. Authors in \cite{Sun:2018IOTJ} overcome this issue by adopting the Q-learning algorithm \cite{Sutton:1998}, which is an iterative algorithm used for characterizing the expected cost for each state-action pair by alternating exploitation and exploration phases. The corresponding offline optimization policy is also characterized in \cite{Sun:2018IOTJ} to be considered as a benchmark for the online algorithm. The authors characterize the privacy-cost trade-off curves, and also evaluate the performance of the proposed algorithm by means of the empirical mutual information. 


\subsection*{Temporal and Spatial Similarities in the Grid Load as a Privacy Measure}

Variations in the grid load profile can be captured by considering power traces of single appliances and by computing differences in power consumption both in the time domain, i.e., consumption deviation over time of a specific appliance, and in the ``space'' domain, i.e., consumption profiles of different appliances. As these variations are computed over a certain time horizon, when an online algorithm is considered, future user electricity consumption is estimated by forecasting the future electricity prices and running Monte Carlo simulations. The optimal decision at any time is characterized by considering both the current inputs and the forecasts through a \textit{rolling online stochastic optimization process}. \textit{Load shifting}, i.e., the scheduling of the user's flexible electricity demand in accordance with privacy as well as cost concerns, can also be considered. Load shifting is analyzed in \cite{Wu:2016}  and in \cite{Chen:2013TSG}, where privacy, cost of energy, and battery wear and tear are jointly optimized, and an online algorithm is formulated. The objective is to minimize the sum of current and expected electricity and charging/discharging costs together with the weighted power profile differences measured through the similarity parameters for an entire day. The effectiveness of three similarity measures are examined in \cite{Chen:2013TSG} separately, as well as jointly, where only four typical appliances, an oven, a clothes dryer, a dishwasher and an electric vehicle, are considered for the sake of simplicity.


\subsection*{Heuristic Algorithms}

While the grid load can be flattened by minimizing the variation of the grid load around a constant consumption target, several works in the literature propose heuristic battery charging and discharging algorithms, which keep the grid load variations limited. An intuitive approach is to try to keep the grid load equal to its most recent value by discharging (charging) the RB when the current user load is larger (smaller) than the previous one. This approach, called the \textit{best-effort} (BE) \textit{algorithm} in \cite{Kalogridis:2010SGC}, tends to eliminate the higher frequency components of the user load, while still revealing the lower frequency components. In \cite{Kalogridis:2010SGC}, the similarity between the two probability distributions of the user and grid load is quantified via the \textit{empirical relative entropy} (\textit{i.e., Kullback-Leibler divergence}) \cite{Cover:1991}. In the same work, the authors also consider \textit{cluster classification}, whereby data is clustered according to power levels, and \textit{cross-correlation} and \textit{regression} procedures, according to which the grid load is shifted in time at the point of maximum cross-correlation with the user load, and regression methods are then used to compare the two aligned signals.


A slightly more sophisticated approach is considered in \cite{McLaughlin:2011}, called the non-intrusive load levelling (NILL) algorithm, in which more than one grid load target value, namely a \textit{steady state target} and \textit{low} and \textit{high recovery state targets} are allowed, and the EMU tries to maintain the grid load at one of these values across time. If the steady state load cannot be maintained, the EMU switches to a high (low) recovery state in case of persistent light (heavy) user demand. When one of the recovery states is reached, the target load is adapted accordingly to permit the battery to charge or discharge, similarly to the empirical strategies outlined in \cite{Yao:2015TSG}. The value of the steady state target load can be updated whenever a recovery state is reached, in order to reduce the occurrences of recovery states, which is achieved by using an exponential weighted moving average of the demand. To assess their proposed approach, the authors in \cite{McLaughlin:2011} count the \textit{number of features}, i.e., the number of times a device is recognized as being on or off, from the grid load, as compared to the user load, and they further consider the \textit{empirical entropy}.

As also pointed out in \cite{Yang:2012}, these heuristic algorithms suffer from precise load change recovery attacks that can identify peaks of user demand. We note that the NILL algorithm is essentially quantizing the input load to three values with the help of the RB. This idea is generalized in \cite{Yang:2012} by considering an arbitrary number of quantization levels. Since quantization is a ``many-to-few mapping'', converting the grid load to a step function is inherently a non-linear and irreversible process, which can be used to provide privacy by maximizing the quantization error under battery limitations. More specifically, the grid load is forced to be a multiple of a quantity $\beta$, i.e., $y_t=h_t \beta$, where $h_t$ is an integer value, and $\beta$ is the largest value that satisfies battery's maximum capacity and power constraints. At any time slot, given the user load, the grid load is chosen between the two adjacent levels to the user load, namely $\ceil[\big]{\frac{x_t}{\beta}}$ and $\floor[\big]{\frac{x_t}{\beta}}$, where $\ceil[\big]{\cdot}$ and $\floor[\big]{\cdot}$ denote the ceiling and floor functions, respectively. Three \textit{stepping algorithms} are proposed in \cite{Yang:2012}, which have different quantization levels: $1$) the \textit{lazy stepping} algorithm, which tries to maintain the external load constant for as long as possible; $2$) the \textit{lazy charging} algorithm, which keeps charging (discharging) the battery until it is full (empty); and $3$) the \textit{random charging} algorithm that chooses its actions at random. While the simulation results show that these algorithms outperform the BE and NILL algorithms, with the lazy stepping algorithm typically performing the best, it is hard to make general claims due to the heuristic nature of these algorithms. In fact, these approaches do not provide theoretical guarantees on the level of privacy achieved; thus, they are not able to make any general claim about the strength of the proposed privacy-preserving approaches and their absolute performance. This is an important limitation as consumers would like to know the level of privacy they can achieve, even if it is in statistical terms. Also, because such heuristics are often based on deterministic schemes, they are prone to be easily reverse-engineered.

\section*{Theoretical Guarantees on SM Privacy}

One of the challenges in SM privacy is to provide theoretical assurances and fundamental limits on the information leaked by an SM system, independently of any assumption on the capability of an attacker, or of the particular NILM algorithm employed. This is essential in privacy research as privacy-preserving techniques may perform extremely well against some NILM algorithms and very poorly against others. Moreover, the privacy assurances should not be based on the complexity limitations of a potential attacker, as techniques that are currently thought not to be feasible, may become available to attackers in the future, if computational capabilities improve, or if new methods are developed. Last but not least, establishing a coherent mathematical framework would allow us to compare various SM scenarios and the use of different physical resources, e.g., RBs of various capacities, RESs of various nature, etc., in a rigorous manner. Accordingly, signal processing and information-theoretic tools have been employed in the literature to provide theoretical privacy assurances. We will overview various different statistical measures for privacy, in particular, the \textit{conditional entropy} \cite{Yao:2015TSG}, \textit{Fisher information} (FI) \cite{Farokhi:2017TSG}, or \textit{type II error probability} in detecting user activity \cite{Zuxing:2015ITW}.

In this statistical framework, it is commonly assumed that the statistics of the user load and the RES are stationary over the period of interest, and are known to the EMU. This assumption is reasonable especially if the period of stationarity is sufficiently long for the EMU to observe and learn these statistics \cite{Qian:2011TPS,Leicester:2016IET,Labeeuw:2013TII}. On the other hand, an online learning theoretic framework can also be considered to account for the convergence time of the learning algorithm. Alternatively, most of the works in the literature that carry out a theoretical analysis also propose suboptimal policies that can be applied on real power traces, thus allowing the reader to gain intuition about the practical application and performance of these theoretically-motivated techniques. We take a worst-case approach, and assume that the statistics  governing the involved random processes are known also by the attacker. Note that this can only empower the attacker, and strengthen the stated privacy guarantees.

\subsection*{The Significance of Single-Letter Expressions}

It is expected that a meaningful privacy measure should consider the leakage of a user's information over a certain time period of reasonable length, because of the memory effects introduced by the RB and the RES. The energy consumption over a short period of time can be easily covered by satisfying all the demand from the RB or the RES over this period, but this may come at the expense of revealing the energy consumption fully at future time periods. Therefore, the information theoretic analysis typically considers an average information rate measured over a given finite time period, and often studies its infinite-horizon asymptotics as well. However, increasing the time horizon also increases the problem complexity, and one of the challenges of the information-theoretic analysis is to obtain a so-called ``single-letter" expression for the optimal solution, which would reduce the problem complexity significantly, particularly when the involved random variables are defined over finite alphabets. Unfortunately, to date, closed-form or single-letter expressions for the information leaked in an SM system have been characterized only for specific settings under various simplifications, e.g., considering an i.i.d. or Markov user load or RES generation.


\subsection*{MI as a Privacy Measure}

The entropy of a random variable $X$, $H(X)$, is a measure of the uncertainty of its realization. The MI between random variables $X$ and $Y$, $I(X;Y)$, measures the amount of information shared between the two random variables \cite{Cover:1991}. MI can also be considered as a measure of dependence between the random variables $X$ and $Y$, and it is equal to zero if and only if they are independent. Rewriting the MI as $I(X;Y)=H(X)-H(X|Y)$, where $H(X|Y)$ is the conditional entropy, we can also interpret MI as the average reduction in the uncertainty of $X$ from the knowledge of $Y$. Therefore, we can measure the privacy leakage about the input load sequence $X^n$ through the SM readings $Y^n$ by the MI between the two sequences, $I(X^n;Y^n)$. This will measure the reduction in the uncertainty of the UP about the real energy consumption of the appliances, $X^n$, after receiving the SM measurements, $Y^n$. For an SM system with only an RB (no RES) and a given EMP $f$ in (\ref{eq:startingPolicy}), running over $n$ time slots, the average \textit{information leakage rate} $\mathcal{I}_f^n (B_{\max},\hat{P}_d)$ is defined as 
\begin{equation}\label{eq:informLeakRate}
\mathcal{I}_f^n (B_{\max},\hat{P}_d) \triangleq \frac{1}{n} I(X^n;Y^n) = \frac{1}{n}\big[ H(X^n)-H(X^n|Y^n) \big],
\end{equation}
where $0 \leq X_t - Y_t \leq \hat{P}_d$. The parameters $B_{\max}$ and $\hat{P}_d$ emphasize the dependence of the EMP, and therefore, of the achievable information leakage rate, on the battery capacity and the discharging peak power constraint. The optimal EMP and the corresponding minimum information leakage rate is obtained by minimizing (\ref{eq:informLeakRate}) over all feasible policies $f \in \mathcal{F}$ to obtain $\mathcal{I}^{n}(B_{\max},\hat{P}_d)$.
\subsubsection*{\textbf{Privacy with an RES}}\label{sec:privacyAES}

Alternatively, one can also consider the SM system of Fig. \ref{fig:systemModel} with an RES, but no RB. Assume that the renewable energy that can be used over the operation period is constrained by an average and a peak power constraint. We do not allow selling the generated renewable energy to the UP, as our goal is to understand the impact of the RES on providing privacy to the user. The minimum information leakage rate achieved under these assumptions and for an i.i.d. user load can be characterized by the so-called \textit{privacy-power function} $\mathcal{I}(\bar{P},\hat{P}_d)$, and can be formulated in the following single-letter form:
\begin{equation}\label{eq:expr_privacy_power}
\mathcal{I}(\bar{P},\hat{P}) = \inf_{p_{Y|X} \in \mathcal{P}} I\left(X;Y\right),
\end{equation}
where $\mathcal{P} \triangleq \{ p_{Y|X}: y \in \mathcal{Y}, \mathbb{E}[(X-Y)] \leq \bar{P}, 0 \leq X-Y \leq \hat{P}\}$. This formulation is presented in \cite{Gunduz:2013ICC} for a discrete user load alphabet, i.e., $X$ can only assume values that are multiples of a fixed quantum, and in \cite{Gomez:2013ISIT}, for a continuous user load alphabet, i.e., $X$ can assume any real value within the limits specified by the peak power constraints of the appliances. The optimal EMP that minimizes (\ref{eq:expr_privacy_power}) is stochastic and memoryless, that is, the optimal grid load at each time slot is generated randomly via the optimal conditional probability that  minimizes (\ref{eq:expr_privacy_power}) by only considering the current user load. Another interesting observation is that Eq. (\ref{eq:expr_privacy_power}) is in a similar form to the well-known \textit{rate-distortion function} in information theory, which characterizes the minimum compression rate $R$ of data, in bits per sample, that is required for the receiver to reconstruct the source sequence within a specified average distortion level $D$ \cite{Cover:1991}. Formally, the rate-distortion function $R(D)$ for an i.i.d. source $X \in \mathcal{X}$ with distribution $p_X$, reconstruction alphabet $\hat{\mathcal{X}}$, and distortion function $d(\hat{x},x)$, where the distortion between sequences $X^n$ and $\hat{X}^n$ is given by $\frac{1}{n}\sum_{i=1}^n d(x_i,\hat{x}_i)$, characterizes the minimum rate with which an average distortion of $D$ is achievable. The compression rate specifies the size of the codebook $2^{nR}$, required to compress the source sequence of length $n$, $X^n$. Shannon showed that the rate-distortion function can be obtained in the following single-letter form:
\begin{equation}\label{eq:rateDistortion}
R(D) = \min_{p_{\hat{X}|X}: \sum_{(x,\hat{x})}p_X p_{\hat{X}|X}d(x,\hat{x})\leq D}  I(\hat{X};X).
\end{equation} 

The analogy between (\ref{eq:expr_privacy_power}) and (\ref{eq:rateDistortion}) becomes clear considering the following distortion measure 
\begin{equation}
d(x,y)=
\begin{cases}
x-y,      &\text{if } 0 \leq x-y \leq \hat{P},\\
\infty,   &\text{otherwise},
\end{cases}
\end{equation}
and such analogy enables using tools from rate-distortion theory to evaluate the privacy-power function for an SM system. However, it is important to highlight that despite the functional similarity, there are major conceptual differences between the two problems, namely: $i$) in the SM privacy problem $Y^n$ is the direct output of the encoder rather than the reconstruction at the decoder side; $ii$) unlike the lossy source encoder, the EMU does not operate over blocks of user load realizations; instead, it operates symbol by symbol, acting instantaneously after receiving the appliance load at each time slot. For discrete user load alphabets, the grid load alphabet can be constrained to the user load alphabet without loss of optimality \cite{Gomez:2015TIFS}, and since MI is a convex function of the conditional probability $p_{Y|X} \in \mathcal{P}$, the privacy-power function can be written as a convex optimization problem with linear constraints. Algorithms such as the Blahut Arimoto (BA) algorithm can be used to numerically compute the optimal conditional distribution \cite{Cover:1991}. For continuous user load distributions, the Shannon lower bound is derived in \cite{Gomez:2015TIFS}, which is a computable lower bound to the rate-distortion function widely used in the literature, and is shown to be tight for exponential user load distributions. 


\begin{figure}
\centering
\includegraphics[width=1\columnwidth]{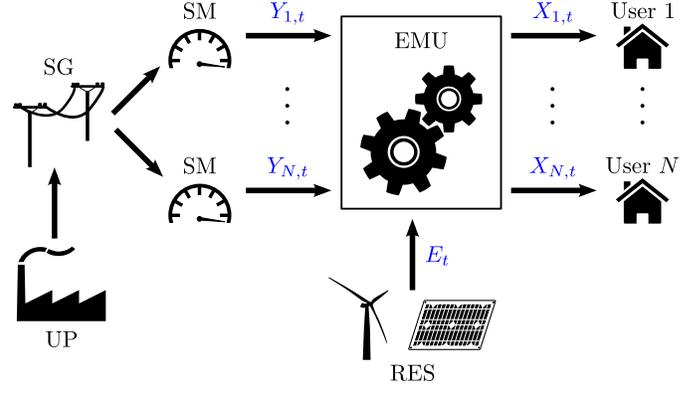}
\caption{A single EMU and RES are shared among $N$ users, each equipped with an SM. The EMU decides how much energy each user can retrieve from the RES and from the grid.}
\label{fig:AESmultipleUsers}
\end{figure}

These results can be generalized to a multi-user scenario in which $N$ users, each equipped with a single SM, share the same RES \cite{Gomez:2015TIFS}. This scenario is represented in Fig. \ref{fig:AESmultipleUsers}, where the objective is to minimize the total privacy loss of $N$ consumers (or, devices) considered jointly, rather than minimizing the privacy loss for each of them separately. This requires the EMU to allocate the shared RES among all the users in the most effective manner. The average information leakage rate can still be written as in (\ref{eq:informLeakRate}), by replacing $X_t$ and $Y_t$ with $\mathbf{X}_t=[X_{1,t},\ldots,X_{N,t}]$ and $\mathbf{Y}_t=[Y_{1,t},\ldots,Y_{N,t}]$, where the boldface characters denote the vectors representing the $N$ power measurements. The privacy-power function has the same expression in (\ref{eq:expr_privacy_power}), and, for the case of independent, but not necessarily identically distributed user loads, the optimization problem (ignoring the peak power constraint) can be cast as 
\begin{equation}\label{expr_privacy_powerMultiple}
\mathcal{I}(\bar{P}) = \inf_{\sum_{i=1}^N P_i\leq \bar{P} } \sum_{i=1}^N \mathcal{I}_{X_i} (P_i),
\end{equation}
where $\mathcal{I}_{X_i} (\cdot)$ denotes the privacy-power function for the $i$-th user having user load distribution $p_{X_i}(x_i)$. For continuous and exponential user loads, the optimal allocation of the energy generated by an RES can be obtained by the \textit{reverse water-filling} algorithm, according to which energy from the RES is only used to satisfy the users with a low average load, while users with higher average load need to request energy from the grid as well.

\subsubsection*{\textbf{Privacy with an RB}}

We can also consider the presence of only an RB in the system, which is thus charged only via the grid (no RES is available to the EMU). Including an RB complicates the problem significantly, and the battery SoC, $B_t$, plays an important role when designing a feasible EMP. 

This problem can be solved by putting it in the form of an MDP and by finding a suitable additive formulation for the privacy cost function \cite{Li:2017Arxiv}. The optimization problem is formulated as
\begin{equation}\label{eq:Li}
L^* \triangleq \min_{f} \frac{1}{n} I(B_{1},X^n;Y^n),
\end{equation}
where $f$ can be any feasible policy, as specified in (\ref{eq:startingPolicy}) (without including the renewable energy process). Eq. (\ref{eq:Li}) has been cast in an additive formulation in \cite{Li:2017Arxiv} by noting that there is no loss of optimality in restricting the focus to charging strategies $f'$ that decide on the grid load only based on the current values of the user load $X_t$ and battery SoC $B_t$, and on the past values of the grid load $Y^{t-1}$, i.e., the general strategy $f$ in (\ref{eq:startingPolicy}) is specified as $f'_t: \mathcal{X} \times \mathcal{B} \times \mathcal{Y}^{t-1} \rightarrow \mathcal{Y}, \forall t$, because of the following inequality:
\begin{equation}\label{eq:specificStrategy}
\frac{1}{n} I(X^n, B_1;Y^n) \geq  \frac{1}{n} \sum_{t=1}^{n} I(X_t,B_t;Y_t|Y^{t-1}).
\end{equation}


The conditional distributions in (\ref{eq:specificStrategy}) grow exponentially with time because of the term $Y^{t-1}$, so that the problem becomes computationally infeasible very quickly. To overcome this problem, the knowledge of $Y^{t-1}$ is summarized into a \textit{belief} state, defined as $p(X_t,B_t|Y^{t-1})$, which can be computed recursively and interpreted as the belief that the UP has about $(X_t,B_t)$ at time $t$, given its past observations, $Y^{t-1}$. This way, the optimal Bellman equations can be formulated, and the optimal policy can be identified numerically (with a discretization of the belief state). 

For an i.i.d. user load, the single-letter characterization of the minimum information leakage rate is given by \cite{Li:2017Arxiv} as
\begin{equation}\label{eq:iidCase}
J^{*} \triangleq \min_{\theta \in \mathcal{P}_B} I(B-X;X),
\end{equation}
where $\theta$ is the probability distribution over $B$ given the past output and actions, i.e., $\theta_t \triangleq p(b_t|y^{t-1},a^{t-1})$, and the action $a_t$ is defined as the transition probability from the current belief, user load and battery SoC to the current grid load. This result is obtained by considering a belief on $W_t \triangleq B_t-X_t$, rather than $(B_t,X_t)$, and by further restricting to policies of the type $f''_t: \mathcal{W} \times \mathcal{Y}^{t-1} \rightarrow \mathcal{Y}, \forall t$. Since (\ref{eq:iidCase}) is convex in $\theta$, the optimal $\theta^*$ may be obtained by using the BA algorithm. The optimal grid load turns out to be i.i.d. and indistinguishable from the demand, while the optimal policy is memoryless and the distribution of $Y_t$ depends only on $W_t$. Such a characterization is provided in \cite{Li:2015SPAWC} for a binary i.i.d. user load, while the authors extend it to an i.i.d. user load of generic alphabet size in \cite{Li:2016Zurich, Li:2016AmerControlConf, Li:2017Arxiv}. 

Another approach is to model the SoC of the RB as a \textit{trapdoor channel} \cite{Permuter:2008TIT}. In a trapdoor channel, a certain number of red or blue balls are within the channel, and a new ball of either color is inserted to it as the channel input at each time step. After the new ball is inserted, one of the balls present in the channel is randomly selected and removed from the channel. In a SM setting, the finite-capacity RB can be viewed as a trapdoor channel, whereby inserting or extracting a ball from the channel represents charging or discharging the RB, respectively. An upper bound on the information leakage rate is characterized in \cite{Arrieta:2017SGC} through this model, by minimizing the information leakage rate over the set of \textit{stable output balls}, i.e., the set of feasible output sequences $Y^n$ that can be extracted from the channel given a certain initial state and an input sequence $X^n$, and by taking inspiration from codebook construction strategies in \cite{Ahlswede:1987TIT}.  The information leakage rate is characterized in \cite{Arrieta:2017SGC} as
\begin{equation}
\frac{1}{n}I(X^n;Y^n) = \frac{1}{\ceil{(B_{\max}+1)/ X_{\max}}},
\end{equation}
where $X_{\max}$ is the largest value $X$ can assume. It is also shown in \cite{Arrieta:2017SGC} that the average user energy consumption determines the level of achievable privacy.


Apart from only maximizing privacy, it is of interest to also minimize the cost. Differently from privacy, cost of energy has an immediate additive formulation and can be easily incorporated into the MDP formulation. Considering the random price vector $C^t=(C_1,\ldots,C_t)$, where $C_t$ denotes the unit cost of energy at time slot $t$, privacy can be defined in the long time horizon as  
\begin{equation}
\mathcal{P} \triangleq \lim_{t \rightarrow \infty} \frac{H(X^t|Y^t,C^t)}{t}.
\end{equation}

This formulation is presented in \cite{Yao:2015TSG}, where the corresponding MDP is formulated, and two suboptimal algorithms are proposed. The first is a \textit{greedy algorithm}, which maximizes at any time the current instantaneous reward, while the second is a \textit{battery centering approach} that is aimed at keeping the battery at a medium level of charge so that the EMU is less constrained by the battery or the demand in determining the grid load. In the latter approach, if the grid load depends not on the current user load or the battery level, but only on the current electricity price, the system is said to be in a \textit{hidden state}, while it is said to be in a \textit{revealing state}, otherwise. The latter strategy is analyzed for an i.i.d. user load by considering the system as a recurrent Markov chain and by adopting random walk theory. 

\subsubsection*{\textbf{Privacy with both an RES and an RB}}

When both an RES and an RB are present, the information theoretic privacy analysis becomes more challenging. As an initial step, we can consider infinite and zero battery capacities, which represent, respectively, lower and upper bounds on the privacy leakage achievable for a practical SM system with a finite-capacity battery \cite{Giaconi:2015,Giaconi:2017TIFS}. When $B_{\max}=\infty$, the problem can be shown to be equivalent to the average and peak power-constrained scenario, and, interestingly, the privacy performance does not deteriorate even if the UP knows the exact amount of renewable energy generated. This shows that, keeping the renewable energy generation process private is more critical when the RB has a limited capacity. Two different energy management policies are shown to achieve the lower bound in\cite{Giaconi:2017TIFS}. In the \textit{best-effort} policy, at any time slot, the optimal EMP derived from (\ref{expr_privacy_powerMultiple}) is employed independently of the RB SoC if there is sufficient energy in the RB, while all the energy request is satisfied from the grid otherwise. The latter one leads to full leakage of user consumption, but it can be shown that these events are rare enough that the information leakage rate does not increase. In the alternative \textit{store-and-hide} policy, an initial storage phase is employed, during which all the energy requests of the user are satisfied from  the  grid  while  all  the  generated  renewable energy is stored in the battery. In the following \textit{hiding} phase, the EMU deploys the optimal policy designed under average and peak power constraints. 

\begin{figure}[!t]
\centering
\includegraphics[width=0.85\columnwidth]{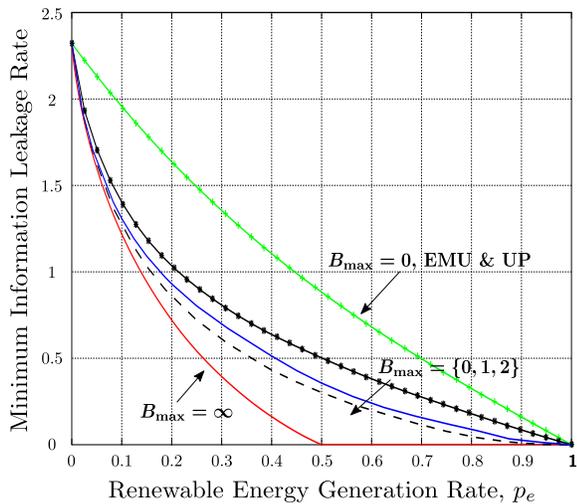}
\caption{Minimum information leakage rate with respect to the renewable energy generation rate $p_e$ with $\mathcal{X}=\mathcal{E}=\mathcal{Y}=\{0,1,2,3,4\}$. The leakage for $B_{\max}=\infty$ has been found by setting $\hat{P}=4$ \cite{Giaconi:2017TIFS}.}
\label{fig:comparison}
\end{figure}

On the other extreme, when $B_{\max}=0$, the renewable energy that can be used at any time slot is limited by the amount of energy generated within that time slot. As expected, assuming the knowledge of the renewable energy process at the UP  significantly degrades the privacy performance for this scenario. Fig. \ref{fig:comparison} compares the minimum information leakage rate with respect to the renewable energy generation rate $p_e$ for  $|\mathcal{X}|=|\mathcal{E}|=|\mathcal{Y}|=5$ when $B_{\max}=\{0,1,2,\infty\}$. In this figure, the curves for a finite battery capacity of $B_{\max}=1$ and $B_{\max}=2$ are obtained numerically by considering a suboptimal EMP \cite{Giaconi:2017TIFS}.

The presence of a finite-capacity battery increases the problem complexity dramatically due to the memory effects induced by the finite battery, and single-letter expressions are still lacking for this scenario. A possible approach to find a theoretical solution to this problem is by extending the MDP formulation, as investigated in \cite{Giaconi:2016}. 

\subsection*{Detection Error Probability as a Privacy Measure}

So far we have considered approaches that try to hide the complete user energy demand from the UP. However, rather than hiding the entire energy consumption profile, in some cases it may be more meaningful to keep private specific user activities, such as ``is there anybody at home?", ``has the alarm been activated?" or ``are you eating microwaved food?". In order to keep the answer to such questions private, the goal of the EMU is to maximize the attacker's probability of making errors when attempting to answer them. 

Let the consumer's behavior that needs to be kept private belong to  a set of $M$ possible activities. Thus, we can treat the attacker's decision and the user's action as an $M$-ary hypothesis, i.e., $H \in \mathcal{H} = \{h_0, h_1, \ldots h_{M-1}\}$. When $M=2$, the hypothesis test is said to be binary and, by convention, the hypothesis $h_0$, called the \textit{null hypothesis}, represents the absence of some factor or condition, while the hypothesis $h_1$, called the \textit{alternative hypothesis}, is the complementary condition. For example, answering the question ``is somebody at home?" corresponds to a binary hypothesis test, where $h_0$ is the hypothesis ``somebody is not at home" and $h_1$ is the hypothesis ``somebody is at home". It is reasonable to assume that the input load will have different statistics under these two hypotheses; accordingly, we assume that under hypothesis $h_0$ ($h_1$), the energy demand at time slot $t$ is i.i.d. with $p_{X|h_0}$ ($p_{X|h_1}$). Based on the SM readings, the attacker  aims at determining the best \textit{decision rule} $\hat{H}(\cdot)$, i.e., the optimal map between the SM readings and the underlying hypothesis. In other words, the space of all possible SM readings, $\mathcal{Y}^n$, is partitioned into the two disjoint decision regions $\mathcal{A}_0$ and $\mathcal{A}_1$, defined as follows:
\begin{align}
\mathcal{A}_0 &\triangleq \{y^n | \hat{H}(y^n) = h_0\},\\
\mathcal{A}_1 &\triangleq \{y^n | \hat{H}(y^n) = h_1\},
\end{align}
which correspond to the subsets of the SM readings for which the UP decides for one of the two hypotheses. The attacker's binary hypothesis test can incur two types of errors:
\begin{itemize}
\item Type I error probability: make a decision $h_1$ when $h_0$ is the true hypothesis  (\textit{false positive} or \textit{false alarm}), i.e., $p_{\mathrm{I}}=p_{Y^n|h_1}(\mathcal{A}_0)$;
\item Type II error probability: make a decision $h_0$ when $h_1$ is the true hypothesis (\textit{false negative} or \textit{miss}), i.e., $p_{\mathrm{II}}=p_{Y^n|h_0}(\mathcal{A}_1)$.
\end{itemize}

The Neyman-Pearson test minimizes the type II error probability for a fixed maximum type I error probability and makes decisions by thresholding the likelihood ratio $\frac{p_{Y^n|h_0}(y^n|h_0)}{p_{Y^n|h_1}(y^n|h_1)}$. Consider the worst case of an all-powerful attacker, which has the perfect knowledge of the EMP employed, in the asymptotic regime $n \rightarrow \infty$, and denote by $p_{\mathrm{II}}^{\min}$ the minimal type II probability of error subject to a constraint on the type I probability of error. Assuming that a memoryless EMP is employed by the EMU, that is, the grid load at any time slot $t$ depends only on the input load at the same time slot, then the attacker runs a Neyman-Pearson detection test on the grid load. We note that the memoryless EMP assumption is not without loss of optimality. However, it is justified on the grounds that, characterizing the more general optimal policy with memory seems to be significantly more challenging, and unlikely to lend itself to a single-letter expression. Chernoff-Stein Lemma \cite{Cover:1991} links the minimal type II error probability $p_{\mathrm{II}}^{\min}$ to the Kullback-Leibler (KL) divergence $D(\cdot||\cdot)$ between the grid load distributions conditioned on the two hypotheses in the limit of the number of observations going to infinity:
\begin{equation} \label{eq:hypotTest}
\lim_{n\rightarrow \infty} - \frac{\log p_{\mathrm{II}}^{\min}}{n} = D(p_{Y|h_0}||p_{Y|h_1}),
\end{equation}
while the KL divergence between two probability distribution functions on $X$, $p_X$ and $q_X$, is defined as \cite{Cover:1991}
\begin{equation}
D(p_X||q_X) \triangleq \sum_{x \in \mathcal{X}} p_X(x) \log\frac{p_X(x)}{q_X(x)}.
\end{equation}

Not surprisingly, to maximize the privacy the goal of the EMU is to find the optimal grid load distributions, which, given the user load $X$ and the true hypothesis $H$, minimizes the KL divergence in (\ref{eq:hypotTest}), or equivalently,  minimizes the asymptotic exponential decay rate of $p_{\mathrm{II}}^{\min}$. However, the EMU is constrained by the available resources in making the two input load distributions produce similar grid load distributions. In particular, we impose a constraint on the average RES it can use. Thus, the objective is to solve the following minimization problem:
\begin{equation}
\min_{p_{Y|H}\in \mathcal{P}_{Y|H}} D(p_{Y|h_0}||p_{Y|h_1}),
\end{equation}
where $\mathcal{P}_{Y|H}$ is the set of feasible energy management policies, i.e., those that satisfy the average RES generation rate $\bar{P}$, so that $\frac{1}{n}\mathbb{E}[\sum_{i=1}^n X_i - Y_i|h_j]\leq \bar{P}$, $j=0,1$. This setting is studied in \cite{Zuxing:2017ISIT}, where the asymptotic single-letter expressions of two privacy-preserving EMPs in the worst case scenario are considered, i.e., when the probability of type $\mathrm{I}$ error is close to $1$. The first policy is a memoryless hypothesis-aware policy that decides on $Y_t$ based only on the current $X_t$ and $H$, while the second policy is unaware of the correct hypothesis $H$ but takes into account all the previous realizations of $X$ and $Y$. It is noteworthy that even if the hypothesis-unaware policy with memory does not have access to the current hypothesis, it performs at least as well as the memoryless hypothesis-aware policy. This is because the hypothesis-unaware policy is able to learn the hypothesis with negligible error probability after observing the energy demand process for a sufficiently long period. Additionally, the energy supply alphabet can be constrained to the energy demand alphabet without loss of optimality, which greatly simplifies the numerical solution to the problem. 

\subsection*{FI as a Privacy Measure}

FI is another statistical measure that can be employed as a measure of SM privacy \cite{Farokhi:2017TSG}. Let some sample data $x$ be drawn according to a distribution depending on an underlying parameter $\theta$. Then, FI is a measure of the amount of information that $x$ contains about $\theta$. In the SM setting, $Y^n$ is the sample data available to the attacker, while $X^n$ is the parameter underlying the sample data that is to be estimated by the UP. Let $\hat{X}^n$ denote the estimate of the UP. The FI can be generalized to the multivariate case by the FI matrix, defined as
\begin{multline}\label{eq:FI}
\mathcal{FI}(x^n) =\\\int_{y^n\in \mathcal{Y}^n} p(y^n|x^n) \bigg[\frac{\partial \log(p(y^n|x^n))}{\partial x^n}\bigg] \bigg[\frac{\partial \log(p(y^n|x^n))}{\partial x^n}\bigg]^T \mathrm{d} y^n.
\end{multline}

Assuming an \textit{unbiased estimator} at the attacker, i.e., the difference between the estimator's expected value and the true average value of the parameter being estimated is zero, the variance of the estimation error can be bounded via the Cram\'er-Rao bound as follows:
\begin{equation}
\mathbb{E}[||x^n-\hat{x}^n(y^n)||_2^2] \geq \mathrm{Tr}(\mathcal{FI}(x^n)^{-1}),
\end{equation}
where $||x^n-\hat{x}^n(y^n)||_2^2$ denotes the squared Euclidean norm, and $\mathrm{Tr}(A)$ denotes the trace of matrix $A$. To maximize the privacy it is then necessary to maximize the trace of the inverse of the FI matrix. In \cite{Farokhi:2017TSG}, two SM settings with RB are studied, specifically when the battery charging policy is independent of the user load, and when it is dependent non-causally on the entire user load sequence. For both cases single-letter expressions are obtained for the maximum privacy. Moreover, the case of biased estimators, wear and tear of the batteries, and peak power charging and discharging constraints are also briefly analyzed in \cite{Farokhi:2017TSG}.


\subsection*{Empirical MI as a  Privacy Measure}

Approaches aimed at determining theoretical privacy limits provide important insights and intuitions for the optimal energy management policy in order to limit the privacy leakage. However, they are often difficult to optimize or even evaluate numerically, and the relatively simplified formulation obtained in various special cases rely on restrictive assumptions, e.g., i.i.d. user load, infinite RB capacity, etc. An alternative is to follow a suboptimal or heuristic EMP. Although such a policy does not provide theoretical privacy guarantees, one can evaluate the corresponding privacy leakage numerically using empirical MI. 

\begin{figure}[!t]
\centering
\includegraphics[width=1\columnwidth]{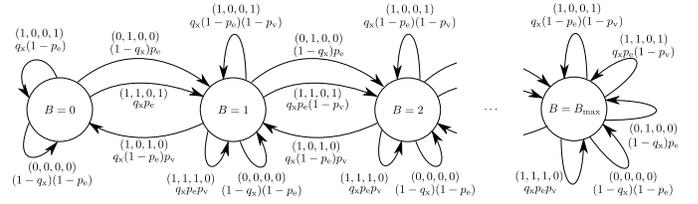}
\caption{An example of RB evolution modelled as an FSM, with $\mathcal{B}=\{0,1,\ldots B_{\max}\}$ and $\mathcal{X}=\mathcal{E}=\mathcal{Y}=\{0,1\}$. The $4$-tuple $(x,e,v,y)$ represent for every time $t$ the values of the user load, the renewable energy produced, the energy taken out of the battery by the EMU, and the grid load, respectively.}
\label{fig:fsm}
\end{figure}

One way to compute the empirical MI is by simulating a discrete time system for a ``large enough" time interval and sampling the resulting $X^n$ and $Y^n$ sequences \cite{Arnold:2006}. The MI between two sequences $x^n$ and $y^n$ can be approximated as
\begin{equation}\label{eq:arnold}
I(X;Y) \approx - \frac{1}{n}\log p(y^n) - \frac{1}{n}\log p(x^n) + \frac{1}{n}\log p(x^n, y^n),
\end{equation}
where $p(y^n)$, $p(x^n)$ and  $p(x^n, y^n)$ are calculated recursively through a sum-product computation. When using this method, the RB is modeled as a finite state machine (FSM), and the battery SoC evolves in time through a Markov chain with transition probabilities depending on the specific policy implemented. An example FSM is illustrated in Fig. \ref{fig:fsm}, where all the processes are considered to be binary and Bernoulli distributed, and the parameters are $q_x = \Pr\{X=1\}$, $p_e = \Pr\{E=1\}$ and $p_v$, the latter being the probability of using energy from the battery provided there is available energy. The support space for the parameters is discretized, and the optimal combination of parameters is found, which minimizes the empirical MI. This approach is followed in \cite{Varodayan:2011ICASSP}, where only a binary RB is present and for an i.i.d. Bernoulli distributed user demand, and in \cite{Tan:2013JSAC} where an RES is also considered. The latter work also analyzes the wasted energy and characterize the privacy-vs-energy efficiency trade-off for the binary scenario and equiprobable user load and renewable energy generation processes. For larger battery capacities and for an equiprobable user load, they note that there is a symmetry and complementarity in the optimal transition probabilities in the FSM model, which simplifies the numerical analysis. This model is also employed in \cite{Giaconi:2015} and \cite{Giaconi:2017TIFS} by considering an RES and designing a suboptimal policy, which, at each time instant, decides among using all of the available energy, half of it, or no energy at all, according to a probability chosen to minimize the overall information leakage.



Another technique for approximating MI is to assume $X$ and $Y$ to be i.i.d.  over a time interval, and approximate the MI via the relative frequency of events $(X_t,Y_t)$ during the same time window. In \cite{Chin:2017TSG} this approach is enriched by additive smoothing, i.e., avoiding zero probability estimates by adding a positive scalar, and it is employed together with a \textit{model-distribution predictive controller}, such that, at each time slot $t$, the EMU decides its actions for a prediction horizon of length $T$, i.e., up to time $t+T$. Privacy and cost are jointly optimized by considering non-causal knowledge of the renewable energy generation process, user load and energy prices, while EMU's actions, i.e., the energy that is requested from the grid and the battery, are forecast over the prediction horizon. The user and grid load processes are assumed to be i.i.d. within a time window $N \gg T$, which also includes the prediction horizon $T$, and finite alphabets $\mathcal{X}$ and $\mathcal{Y}$ are considered. As $N \gg T$, first-order Taylor approximation of the logarithm function is used, and the corresponding mixed integer quadratic program is formulated, which is of manageable size and can be solved recursively whenever new SM readings are available. Results show that considering a relatively small prediction horizon $T$ prevents the EMU from fully utilizing the RB capacity, as the user load that is considered by the algorithm is generally smaller then the RB capacity. Allowing a longer prediction interval dramatically improves the performance in terms of both privacy and cost, at the expense of a much higher computational complexity. The work also shows that by increasing the alphabet sizes of $\mathcal{X}$ and $\mathcal{Y}$ better privacy performance can be achieved. 

Empirical MI normalized by the empirical entropy of the user load is considered in \cite{Koo:2012}, where an RB is used to minimize the energy cost subject to privacy constraints. Here two cost tariffs are considered, a low-price and a high-price, and a dynamic programming approach is developed to maximize the energy stored in the battery at the end of the low-price period, and minimize it at the end of the high-price period. At every time slot, the optimal probability distribution of the grid load is computed, which is forced to be independent of the user load distribution.





\section*{Concluding Remarks and Future Challenges} \label{sec:conclusion}

Privacy and ``the right to be let alone" are an individual's inalienable fundamental rights, which are safeguarded in many national constitutions worldwide. In Europe, the General Data Protection Regulation (GDPR) \cite{GDPR:2016}, which will be enforced from 25 May 2018, will set even more stringent requirements for every technology or device that collects and processes customer data, including SMs. For these reasons, addressing the SM privacy problem is crucial for the adoption of the SG concept. In fact, considering the growing privacy concerns of the consumers for SMs as well as many other emerging technologies \cite{pwc:2017}, a critical growth in SM adoption and other SG technologies will take place only when consumers are given full control of their privacy, and they feel they have clear and honest information on how their data is being used. Only then, consumer resistance can be overcome and their trust will be assured, thus paving the way to a more fertile and fair ground for new products and increased innovation in this domain. 

UPs and their partners, including governments, may be too keen on collecting users' data indiscriminately, and less incentivized to develop privacy-enhancing technologies. Therefore, legislators, public commissions, consumer advocacy groups and researchers have an important role in tackling the SM privacy problem, and preventing the SM data from being gathered indiscriminately and sold to third parties without explicit user consent, or even passed to government intelligence agencies for mass surveillance scopes. GDPR is a good example of such initiatives. However, given that such a legal framework is still lacking and not yet fully developed globally, it becomes imperative to push forward the concept of \textit{privacy-by-design}, according to which privacy should be ``designed-in" to new products and services, rather than considered only after user complaints and regulatory impositions. This is because a wider range of options are available during the design stage as compared to modifying the product following a privacy incident or a user complaint. Achieving privacy-by-design is the ultimate goal of the techniques analyzed in this paper. 

In this article we have focused exclusively on techniques that adopt physical resources, such as RESs and RBs, to provide privacy to users. Main motivation and benefits of these techniques is that they do not undermine the benefits of the SG concept. Each of the outlined techniques has its unique advantages and disadvantages, and focuses on a particular aspect of privacy. However, despite the considerable efforts put into developing SM privacy-preserving techniques, the full extent of the privacy problem in SMs is far from being completely understood, and a unified and coherent vision for SM privacy (just like in many other domains) is still missing.  

In the context of SM privacy, UDS-based methods manipulate a physical quantity, energy, to ensure privacy for users. This entails that physical constraints, such as those related to an RB or an RES, play a crucial role in finding the optimal privacy-preserving strategy. We expect that the techniques developed for enhancing SM privacy can prove useful in other privacy-sensitive settings, in which physical quantities are involved, such as gas and water meters, or location privacy.

\subsection*{Research Challenges}

Various challenges must be addressed before privacy-by-design can become a reality in SM systems. Firstly, a generic privacy measure, or a combination of different measures, must be determined and adopted in order to formally quantify the loss of privacy, in the same way a user's electricity bill is computed. Such a measure should be device-independent, and enable the comparison of various privacy-preserving strategies. It is also necessary to understand the implications of the various privacy measures on the grid load. From this point of view, theoretical measures may be preferable due to their abstract and fundamental nature, i.e., they are independent of any assumptions on the attacker's algorithms. However, their relevance in real-world scenarios must be assessed further, and, if necessary, valid suboptimal privacy measures or algorithms should be put forward and standardized as a proxy for more rigorous privacy assurances. 

Another important goal is to give consumers as much flexibility as possible in setting their desired level of privacy, trading off privacy with the cost of electricity, or other services. It is also essential to allow consumers the possibility of setting different privacy requirements for different devices, as users may value the information about the usage of a certain device more sensitive compared to others. This may happen because certain devices are naturally more correlated to the user's activities or presence at home, such as the use of a kettle, a microwave or an oven, or because a user may decide to hide the usage of a certain appliance for personal reasons.

In the near future, a wider use of electric vehicles will also bring additional complications to the SM privacy problem, as mobility patterns may be inferred by analyzing the charging and discharging events. This problem can be tackled by load shifting, which is expected to play an important role in jointly optimizing electricity cost and privacy. Load shifting, as well as other privacy-preserving techniques introduced here, will be more accurate and relevant thanks to the development of reliable prediction techniques for future electricity consumption, e.g., by using machine learning techniques. The proliferation of various energy-hungry ``smart devices'' will complicate the problem further and overburden RBs even more. Finally, the use of shared physical resources should also be investigated in more depth, as cities are becoming more and more dense and users may want to team up to install storage devices or energy generators that are still rather costly. In cities, solar panels or mini wind turbines may be installed on the roof of blocks of flats and RBs may be put in communal areas, and these resources can be used jointly by all the users in a building. Such resource sharing models make the privacy problem even more complicated and challenging, and might call for a game theoretic formulation of the problem.

Overall, we hope that presenting this overview of the SM privacy problem and current solutions will further encourage research and development in this area, so that remaining open issues will be solved and the SMs' full potential will be unleashed.

\bibliographystyle{IEEEtran}
\bibliography{SPM}

\section*{Authors}
\textit{\textbf{Giulio Giaconi}} (g.giaconi@imperial.ac.uk) received the B.Sc. and M.Sc. degrees (Hons.) in communications engineering from Sapienza University of Rome, Italy, in 2011 and 2013, respectively. He is a Ph.D. candidate with the Department of Electrical and Electronic Engineering, Imperial College London, UK, and a research scientist with BT Research and Innovation, Security Futures Practice. In 2013, he was a Visiting Student with Imperial College London, working on indoor localization via visible light communications. His current research interests include cybersecurity, data privacy, information and communication theory, signal processing and machine learning. In 2014, he received the Excellent Graduate Student Award of the Sapienza University of Rome.

\textit{\textbf{Deniz G\"{u}nd\"{u}z}} (d.gunduz@imperial.ac.uk) received his M.S. and Ph.D. degrees from NYU Polytechnic School of Engineering in 2004 and 2007, respectively. He is a Reader in information theory and communications at the Electrical and Electronic Engineering Department of Imperial College London, UK. He is an Editor of the IEEE Transactions on Communications, and the IEEE Transactions on Green Communications and Networking. He is the recipient of the IEEE Communications Society CTTC Early Achievement Award in 2017, a Starting Grant of the European Research Council (ERC) in 2016, IEEE Communications Society Best Young Researcher Award for the Europe, Middle East, and Africa Region in 2014, and the Best Paper Award at the 2016 IEEE Wireless Communications and Networking Conference (WCNC). 

\textit{\textbf{H. Vincent Poor}} (poor@princeton.edu) is the Michael Henry Strater University Professor of Electrical Engineering at Princeton University. His research interests include information theory and signal processing, and their applications in wireless communications, energy systems and related fields. An IEEE Fellow, he is also a member of the National Academy of Engineering and the National Academy of Sciences, and a foreign member of the Chinese Academy of Sciences, the Royal Society, and other national and international academies.  He received the Technical Achievement and Society Awards of the IEEE Signal Processing Society in 2007 and 2011, respectively. Recent recognition of his work includes the 2017 IEEE Alexander Graham Bell Medal, and a D.Sc. \textit{honoris causa} from Syracuse University, awarded in 2017.

\end{document}